\documentclass[
  a4paper,
  superscriptaddress,
  twocolumn,
  prx,
  preprintnumbers,
  floatfix,
  secnumarabic,
  amssymb
]
{revtex4-2}
\usepackage{lineno}
\usepackage{dcolumn}   
\usepackage{bm}        
\usepackage{amssymb}   
\usepackage{amsmath,stmaryrd}
\usepackage{blkarray, multirow, graphicx, diagbox, color, xcolor, colortbl}
\usepackage{bbm, bbold}
\usepackage{glossaries}
\usepackage{hyphenat}
\usepackage{ifthen}
\usepackage{float}
\usepackage{xkeyval}
\usepackage{moreverb}
\usepackage{rotating}
\usepackage{wrapfig}
\usepackage{slashbox}
\usepackage{xspace}
\usepackage{nicefrac}
\usepackage[]{units}
\usepackage{physics}
\usepackage{booktabs}
\usepackage{braket}
\usepackage{float}
\usepackage[inline]{enumitem}
\usepackage{tabto}
\usepackage{listings}
\usepackage{xstring}
\usepackage{lipsum}
\usepackage{scalerel}

\def\ReplaceStr#1{%
	\IfSubStr{#1}{p}{%
		\StrSubstitute{#1}{p}{.}}{#1}}

\usepackage[caption=false]{subfig}
\newcommand\subfigref[1]{\protect\subref{#1}}

\usepackage[customcolors]{hf-tikz} 
\usepackage{tikz}
\usepackage{calc}
\usetikzlibrary{external}
\usepackage{pgffor}
\usepackage{pgfplots}
\pgfplotsset{compat=1.13}
\usepackage{pgfplotstable}
\usepgfplotslibrary{groupplots}
\usepgfplotslibrary{fillbetween}
\usepgfplotslibrary{colorbrewer}

\tikzstyle{n} = [draw,shape=ellipse,minimum size=1.5em,inner sep=0pt,fill=white!20, minimum width=2.5em]
\tikzstyle{Init} = [n,color=green,fill=green!20,text=black]
\tikzstyle{Fin} = [n,color=red,fill=red!20,text=black]
\tikzstyle{Ghost} = [minimum size=1.5em,inner sep=0pt,color=white,text=black]
\tikzstyle{Multiple} = [draw,shape=rect,minimum size=2em,inner sep=0pt]

\tikzstyle{ghostA} = [text=red!70,thick, minimum size=2*(5pt-\pgflinewidth), inner sep=0pt, outer sep=0pt]
\tikzstyle{ghostB} = [text=blue!70,thick, minimum size=2*(5pt-\pgflinewidth), inner sep=0pt, outer sep=0pt]
\tikzstyle{siteA} = [regular polygon, regular polygon sides=3, shape border rotate= 30, draw=red!50,fill=red!20,thick,inner sep=0pt,minimum width=1.5em,font=\footnotesize]
\tikzstyle{siteB} = [regular polygon, regular polygon sides=3, shape border rotate= -30, draw=green!50,fill=green!20,thick,inner sep=0pt,minimum width=1.5em,font=\footnotesize]
\tikzstyle{op} = [regular polygon, regular polygon sides=4, draw=orange!50, fill=orange!20, thick, inner sep=0.2pt, minimum width=0.25em, minimum height=0.5em,font=\footnotesize]
\tikzstyle{gate} = [rectangle, rounded corners=2pt, draw=orange!50, fill=orange!20, thick, inner sep=0.2pt, minimum width=0.25em, minimum height=0.5em,font=\footnotesize]
\tikzstyle{opghost} = [regular polygon, regular polygon sides=4, thick, inner sep=0.2pt, minimum width=1.25em, minimum height=1.5em,font=\footnotesize]
\tikzstyle{site} = [circle,draw=blue!50,fill=blue!20,thick,inner sep=0.2pt,minimum width=1.25em,font=\footnotesize]
\tikzstyle{hiddensite} = [circle,draw=white!50,fill=white!20,thick,inner sep=0.2pt,minimum width=1.25em,font=\footnotesize]
\tikzstyle{nosite} = [circle,draw=white,fill=white,thick,inner sep=0.1pt,minimum width=1.5em]
\tikzstyle{ghost} = [font=\footnotesize]
\tikzstyle{intersite} = [regular polygon, regular polygon sides=4, shape border rotate= 45, draw=black!50,fill=black!20,thick,inner sep=0pt,minimum width=1.5em]
\tikzstyle{ld} = [inner sep=1pt, font=\small]
\tikzstyle{unsite} = [circle, outer sep=0pt,inner sep=0.2pt,minimum width=1.25em]

\definecolor{colorA}{rgb} {0.58,0,0.8275}
\definecolor{colorB}{rgb} {0.11,0.663,0.51}
\definecolor{colorC}{rgb} {0.3373,0.7059,0.9137}
\definecolor{colorD}{rgb} {0.902,0.6235,0}
\definecolor{colorE}{rgb} {0.9451,0.902,0.3255}
\definecolor{colorF}{rgb} {0.3373,0.3255,0.902}
\definecolor{colorG}{rgb} {0.9451,0.3255,0.3373}
\definecolor{cbColorA}{HTML} {2D4C8D}
\definecolor{cbColorB}{HTML} {EB821D}
\definecolor{cbColorC}{HTML} {C3342F}
\definecolor{cbColorD}{HTML} {24451B}
\definecolor{cbColorE}{HTML} {2C0041}
\definecolor{cbColorF}{HTML} {D9D9D9}

\definecolor{stBlue}{HTML} {5566AA}
\definecolor{stGreen}{HTML} {117733}
\definecolor{stCyan}{HTML} {33BBEE}
\definecolor{stTeal}{HTML} {009988}
\definecolor{stOrange}{HTML} {EE7733}
\definecolor{stYellow}{HTML} {F7F056}
\definecolor{stRed}{HTML} {CC3311}
\definecolor{stMagenta}{HTML} {EE3377}
\definecolor{stGrey}{HTML} {BBBBBB}

\usetikzlibrary
{
	calc,
	decorations,
	plotmarks,
	patterns,
	positioning,
	petri,
	arrows,
	decorations.markings,
	backgrounds,
	fit,
	graphs,
	shapes.geometric,
	decorations.pathmorphing,
	shapes.misc,
	shapes,
	tikzmark,
	pgfplots.colorbrewer,
	fpu,
	svg.path
}

\usepackage{ocgx}

\usetikzlibrary{ocgx}

\pgfplotsset{
        cycle from colormap manual style/.style={
            x=3cm,y=10pt,ytick=\empty,
            stack plots=y,
            every axis plot/.style={line width=2pt},
        },
}

\tikzset{>=stealth}

\tikzset{->-/.style={decoration={
			markings,
			mark=at position .5 with {\arrow{>}}},postaction={decorate}}}

\tikzset{-<-/.style={decoration={
			markings,
			mark=at position .5 with {\arrow{<}}},postaction={decorate}}}

\tikzstyle{orientedsnake} = [
decorate, 
decoration={snake},
->
]  
\tikzstyle{orientedshortarrow} = [
decoration={markings,
	mark=at position .33 with {\arrow{>}}},
postaction={decorate}
]  
\tikzstyle{orientedlongarrow} = [
decoration={markings,
	mark=at position .67 with {\arrow{>}}},
postaction={decorate}
]
\tikzset{dbl/.style={double,
		double equal sign distance,
		-implies,
		shorten >=10pt,
		shorten <=10pt}}
\tikzset{
	between/.style args={#1 and #2}{
		at = ($(#1)!0.5!(#2)$)
	}
}

\tikzstyle{process} = [rectangle, minimum width=3cm, minimum height=1cm, text centered, text width=5cm, draw=black]
\tikzstyle{io} = [trapezium, trapezium left angle=70, trapezium right angle=110, minimum width=3cm, minimum height=1cm, text centered, text width=7cm, draw=black]
\tikzstyle{choose} = [diamond, inner sep=1pt, minimum width=2cm, minimum height=2cm, text centered, text width=1.5cm, draw=black]
\tikzstyle{arrow} =[thick,->, >=stealth]

\pgfmathdeclarefunction{peierlspotential}{6}%
{
	\pgfmathparse
	{
		#2 / (#3 * #5) 
		* sin(deg(#1 * #3 - #3 * #5 * (x)))
		* exp(-( ( #1 - #5 * ((x) - #4) ) )^2.0 / ( #6^2.0 ) )
	}%
}

\pgfmathdeclarefunction{linearFct}{2}%
{%
	\pgfmathparse{#1*x+#2}%
}
\pgfmathdeclarefunction{quadFct}{3}%
{%
	\pgfmathparse{#1*x*x+#2*x+#3}%
}
\pgfmathdeclarefunction{logFct}{2}%
{%
	\pgfmathparse{#1*log10(x)+#2}%
}
\pgfmathdeclarefunction{expFct}{2}%
{%
	\pgfmathparse{#1*exp(-x/#2)}%
}

\newboolean{buildCurrentBlockInline}
\setboolean{buildCurrentBlockInline}{false}
\newboolean{rebuildCurrentBlock}
\setboolean{rebuildCurrentBlock}{false}

\newboolean{rebuildData}
\setboolean{rebuildData}{false}

\newboolean{removeData}
\setboolean{removeData}{false}

\graphicspath{{figures/}}

\newboolean{buildtikzpics}
\setboolean{buildtikzpics}{false}
\newif\ifrebuildtikz
\newif\ifChangeMode
\ChangeModetrue
\ChangeModefalse
\ifthenelse{\boolean{buildtikzpics}}
{
	\rebuildtikztrue
	\usetikzlibrary{external}
	\tikzexternalize[optimize=false,prefix=figures/autogen/]%
}
{
	\rebuildtikzfalse
}

\usepackage[normalem]{ulem}
\usepackage[colorlinks,bookmarks=false,citecolor=blue,linkcolor=black,urlcolor=blue]{hyperref}
\usepackage{cleveref}
\Crefname{appendix}{Appendix}{Appendices}
\Crefname{equation}{Equation}{Equations}
\Crefname{figure}{Figure}{Figures}
\Crefname{section}{Section}{Sections}
\Crefname{tabular}{Tabular}{Tabulars}
\crefname{appendix}{App.}{Apps.}
\crefname{equation}{Eq.}{Eqs.}
\crefname{figure}{Fig.}{Figs.}
\crefname{section}{Sec.}{Secs.}
\crefname{tabular}{Tab.}{Tabs.}

\ExplSyntaxOn
\DeclareExpandableDocumentCommand \eval { m } { \fp_eval:n { #1 } }
\ExplSyntaxOff

\makeatletter
\def\pgfplotsutil@decstringcounter#1{%
 \begingroup
  \c@pgf@counta=#1\relax
  \advance\c@pgf@counta by -1
  \edef#1{\the\c@pgf@counta}%
  \pgfmath@smuggleone#1%
 \endgroup
}%

\pgfplotsset{
/pgfplots/each nth point*/.style 2 args={%
/pgfplots/x filter/.append code={%
 \ifnum\coordindex=0
  \def\c@pgfplots@eachnthpoint@xfilter{0}%
  \edef\c@pgfplots@eachnthpoint@xfilter@cmp{#1}%
 \else
  \ifnum\coordindex>#2\relax
   \pgfplotsutil@advancestringcounter\c@pgfplots@eachnthpoint@xfilter
   \ifx\c@pgfplots@eachnthpoint@xfilter@cmp\c@pgfplots@eachnthpoint@xfilter
    \def\c@pgfplots@eachnthpoint@xfilter{0}%
   \else
    \let\pgfmathresult\pgfutil@empty
   \fi
  \fi
 \fi
}%
},
}
\makeatother

\newcommand{\printpgfnumberorder}[1]%
{%
	\pgfmathfloatparsenumber{#1}%
	\pgfmathfloattomacro{\pgfmathresult}{\Fn}{\Mn}{\En}%
	\pgfmathparse{\Fn==2 ? "-" : ""}%
	\edef\Sn{\pgfmathresult}%
	\Sn 10^{\En}%
}

\newcounter{marknumber}
\pgfplotsset{
	error bars/every nth mark/.style={
		/pgfplots/error bars/draw error bar/.prefix code={
			\pgfmathtruncatemacro\marknumbercheck{mod(floor(\themarknumber/2),#1)}
			\ifnum\marknumbercheck=0
			\else
			\begin{scope}[opacity=0]
				\fi
			},
			/pgfplots/error bars/draw error bar/.append code={
				\ifnum\marknumbercheck=0
				\else
			\end{scope}
			\fi
			\stepcounter{marknumber}    
		}
	}
}

\pgfplotsset{
    /pgf/declare function={
        vk(\x,\a,\b,\c) = \c*sqrt(1.0-(-2*cos(\x)-\b)/sqrt((-2*cos(\x)-\b)^2+\a^2));
    },
}

\pgfplotsset{
/pgfplots/colormap={magma}{
	rgb255=(0,0,4) rgb255=(0,0,6) rgb255=(1,0,7) rgb255=(1,1,9) rgb255=(1,1,11) rgb255=(2,2,13) rgb255=(2,2,15) rgb255=(3,3,17) rgb255=(4,3,19) rgb255=(4,4,21) rgb255=(5,4,23) rgb255=(6,5,25) rgb255=(7,5,27) rgb255=(8,6,29) rgb255=(9,7,31) rgb255=(10,7,34) rgb255=(11,8,36) rgb255=(12,9,38) rgb255=(13,10,40) rgb255=(14,10,42) rgb255=(15,11,44) rgb255=(16,12,47) rgb255=(17,12,49) rgb255=(18,13,51) rgb255=(20,13,53) rgb255=(21,14,56) rgb255=(22,14,58) rgb255=(23,15,60) rgb255=(24,15,63) rgb255=(26,16,65) rgb255=(27,16,68) rgb255=(28,16,70) rgb255=(30,16,73) rgb255=(31,17,75) rgb255=(32,17,77) rgb255=(34,17,80) rgb255=(35,17,82) rgb255=(37,17,85) rgb255=(38,17,87) rgb255=(40,17,89) rgb255=(42,17,92) rgb255=(43,17,94) rgb255=(45,16,96) rgb255=(47,16,98) rgb255=(48,16,101) rgb255=(50,16,103) rgb255=(52,16,104) rgb255=(53,15,106) rgb255=(55,15,108) rgb255=(57,15,110) rgb255=(59,15,111) rgb255=(60,15,113) rgb255=(62,15,114) rgb255=(64,15,115) rgb255=(66,15,116) rgb255=(67,15,117) rgb255=(69,15,118) rgb255=(71,15,119) rgb255=(72,16,120) rgb255=(74,16,121) rgb255=(75,16,121) rgb255=(77,17,122) rgb255=(79,17,123) rgb255=(80,18,123) rgb255=(82,18,124) rgb255=(83,19,124) rgb255=(85,19,125) rgb255=(87,20,125) rgb255=(88,21,126) rgb255=(90,21,126) rgb255=(91,22,126) rgb255=(93,23,126) rgb255=(94,23,127) rgb255=(96,24,127) rgb255=(97,24,127) rgb255=(99,25,127) rgb255=(101,26,128) rgb255=(102,26,128) rgb255=(104,27,128) rgb255=(105,28,128) rgb255=(107,28,128) rgb255=(108,29,128) rgb255=(110,30,129) rgb255=(111,30,129) rgb255=(113,31,129) rgb255=(115,31,129) rgb255=(116,32,129) rgb255=(118,33,129) rgb255=(119,33,129) rgb255=(121,34,129) rgb255=(122,34,129) rgb255=(124,35,129) rgb255=(126,36,129) rgb255=(127,36,129) rgb255=(129,37,129) rgb255=(130,37,129) rgb255=(132,38,129) rgb255=(133,38,129) rgb255=(135,39,129) rgb255=(137,40,129) rgb255=(138,40,129) rgb255=(140,41,128) rgb255=(141,41,128) rgb255=(143,42,128) rgb255=(145,42,128) rgb255=(146,43,128) rgb255=(148,43,128) rgb255=(149,44,128) rgb255=(151,44,127) rgb255=(153,45,127) rgb255=(154,45,127) rgb255=(156,46,127) rgb255=(158,46,126) rgb255=(159,47,126) rgb255=(161,47,126) rgb255=(163,48,126) rgb255=(164,48,125) rgb255=(166,49,125) rgb255=(167,49,125) rgb255=(169,50,124) rgb255=(171,51,124) rgb255=(172,51,123) rgb255=(174,52,123) rgb255=(176,52,123) rgb255=(177,53,122) rgb255=(179,53,122) rgb255=(181,54,121) rgb255=(182,54,121) rgb255=(184,55,120) rgb255=(185,55,120) rgb255=(187,56,119) rgb255=(189,57,119) rgb255=(190,57,118) rgb255=(192,58,117) rgb255=(194,58,117) rgb255=(195,59,116) rgb255=(197,60,116) rgb255=(198,60,115) rgb255=(200,61,114) rgb255=(202,62,114) rgb255=(203,62,113) rgb255=(205,63,112) rgb255=(206,64,112) rgb255=(208,65,111) rgb255=(209,66,110) rgb255=(211,66,109) rgb255=(212,67,109) rgb255=(214,68,108) rgb255=(215,69,107) rgb255=(217,70,106) rgb255=(218,71,105) rgb255=(220,72,105) rgb255=(221,73,104) rgb255=(222,74,103) rgb255=(224,75,102) rgb255=(225,76,102) rgb255=(226,77,101) rgb255=(228,78,100) rgb255=(229,80,99) rgb255=(230,81,98) rgb255=(231,82,98) rgb255=(232,84,97) rgb255=(234,85,96) rgb255=(235,86,96) rgb255=(236,88,95) rgb255=(237,89,95) rgb255=(238,91,94) rgb255=(238,93,93) rgb255=(239,94,93) rgb255=(240,96,93) rgb255=(241,97,92) rgb255=(242,99,92) rgb255=(243,101,92) rgb255=(243,103,91) rgb255=(244,104,91) rgb255=(245,106,91) rgb255=(245,108,91) rgb255=(246,110,91) rgb255=(246,112,91) rgb255=(247,113,91) rgb255=(247,115,92) rgb255=(248,117,92) rgb255=(248,119,92) rgb255=(249,121,92) rgb255=(249,123,93) rgb255=(249,125,93) rgb255=(250,127,94) rgb255=(250,128,94) rgb255=(250,130,95) rgb255=(251,132,96) rgb255=(251,134,96) rgb255=(251,136,97) rgb255=(251,138,98) rgb255=(252,140,99) rgb255=(252,142,99) rgb255=(252,144,100) rgb255=(252,146,101) rgb255=(252,147,102) rgb255=(253,149,103) rgb255=(253,151,104) rgb255=(253,153,105) rgb255=(253,155,106) rgb255=(253,157,107) rgb255=(253,159,108) rgb255=(253,161,110) rgb255=(253,162,111) rgb255=(253,164,112) rgb255=(254,166,113) rgb255=(254,168,115) rgb255=(254,170,116) rgb255=(254,172,117) rgb255=(254,174,118) rgb255=(254,175,120) rgb255=(254,177,121) rgb255=(254,179,123) rgb255=(254,181,124) rgb255=(254,183,125) rgb255=(254,185,127) rgb255=(254,187,128) rgb255=(254,188,130) rgb255=(254,190,131) rgb255=(254,192,133) rgb255=(254,194,134) rgb255=(254,196,136) rgb255=(254,198,137) rgb255=(254,199,139) rgb255=(254,201,141) rgb255=(254,203,142) rgb255=(253,205,144) rgb255=(253,207,146) rgb255=(253,209,147) rgb255=(253,210,149) rgb255=(253,212,151) rgb255=(253,214,152) rgb255=(253,216,154) rgb255=(253,218,156) rgb255=(253,220,157) rgb255=(253,221,159) rgb255=(253,223,161) rgb255=(253,225,163) rgb255=(252,227,165) rgb255=(252,229,166) rgb255=(252,230,168) rgb255=(252,232,170) rgb255=(252,234,172) rgb255=(252,236,174) rgb255=(252,238,176) rgb255=(252,240,177) rgb255=(252,241,179) rgb255=(252,243,181) rgb255=(252,245,183) rgb255=(251,247,185) rgb255=(251,249,187) rgb255=(251,250,189) rgb255=(251,252,191)}
}

    \newacronym[shortplural={MPS}]{MPS}{MPS}{matrix\hyp product state}
\newacronym{MPSs}{MPS}{matrix\hyp product states}
\newacronym[shortplural={PP-MPS}]{PP-MPS}{PP-MPS}{projected purified matrix\hyp product state}
\newacronym{PP-MPSs}{PP-MPS}{projected purified matrix\hyp product states}
\newacronym{MPO}{MPO}{matrix-product operator}
\newacronym{SVD}{SVD}{singular-value decomposition}
\newacronym{QCS}{QCS}{quantum-computer simulator}
\newacronym{FSM}{FSM}{finite-state machine}
\newacronym{ACA}{ACA}{adaptive cross approximation}
\newacronym{1D}{1D}{one\hyp dimensional}
\newacronym{QC}{QC}{quantum computer}
\newacronym{CDW}{CDW}{charge\hyp density wave}
\newacronym{BOW}{BOW}{bond\hyp order wave}
\newacronym{SDW}{SDW}{spin\hyp density wave}
\newacronym{ARPES}{ARPES}{angle-resolved photoemission spectroscopy}
\newacronym{OBC}{OBC}{open-boundary conditions}
\newacronym{PBC}{PBC}{periodic-boundary conditions}
\newacronym{TEBD}{TEBD}{time-evolving block-decimation}
\newacronym{TDVP}{TDVP}{time\hyp dependent variational principle}
\newacronym{iff}{iff}{if and only if}
\newacronym{DFT}{DFT}{density\hyp functional theory}
\newacronym{DMFT}{DMFT}{dynamical mean\hyp field theory}
\newacronym{DMRG}{DMRG}{density\hyp matrix renormalization group}
\newacronym{1DMRG}{1DMRG}{single-site density\hyp matrix renormalization group}
\newacronym{2DMRG}{2DMRG}{two-site density\hyp matrix renormalization group}
\newacronym{DMRG3S}{DMRG3S}{strictly single-site density\hyp matrix renormalization group}
\newacronym{iDMRG}{iDMRG}{inifinite\hyp size density\hyp matrix renormalization group}
\newacronym{tDMRG}{tDMRG}{time\hyp dependent density\hyp matrix renormalization group}
\newacronym{PP-DMRG}{PP-DMRG}{projected purified density\hyp matrix renormalization group}
\newacronym{QMC}{QMC}{quantum Monte Carlo}
\newacronym{AIM}{AIM}{Anderson impurity model}
\newacronym{SIAM}{SIAM}{single impurity Anderson model}
\newacronym{LDA}{LDA}{local\hyp density approximation}
\newacronym{LBNL}{LBNL}{Lawrence Berkeley National Laboratory}
\newacronym{ED}{ED}{exact diagonalization}
\newacronym{QPT}{QPT}{quantum phase transition}
\newacronym{QCP}{QCP}{quantum critical point}
\newacronym{ETH}{ETH}{eigenstate thermalization hypothesis}
\newacronym{EHM}{EHM}{extended Hubbard model}
\newacronym{AKLT}{AKLT}{Affleck\hyp Lieb\hyp Kennedy\hyp Tasaki}
\newglossaryentry{QR}{name={QR},description={QR decomposition}}
\newacronym{TNS}{TNS}{tensor\hyp network state}
\newacronym{SM}{SM}{supplemental material}
\newacronym{NOO}{NOO}{natural orbital occupation}
\newacronym{NO}{NO}{natural orbital}
\newacronym{LRO}{LRO}{long\hyp range order}
\newacronym{qLRO}{qLRO}{quasi\hyp long\hyp range order}
\newacronym{SC}{SC}{Superconductivity}
\newacronym{VBGS}{VBGS}{valence bond ground-state}
\newacronym{PEPS}{PEPS}{projected entangled pair\hyp states}
\newacronym{ALS}{ALS}{alternating least squares}
\newacronym{BdG}{BdG}{Bogoljubov de-Gennes}
\newacronym{TFIM}{TFI}{transverse field Ising model}
\newacronym{PP}{PP}{projected purification}
\newacronym{BEC}{BEC}{Bose\hyp Einstein condensate}
\newacronym{JWT}{JWT}{Jordan\hyp Wigner transformation}
\newacronym{NISQ}{NISQ}{noisy intermediate scale quantum}
\newacronym{NN}{NN}{nearest\hyp neighbor}
\newacronym{NNN}{NNN}{next\hyp nearest\hyp neighbor}
\newacronym{SPDM}{SPDM}{single\hyp particle density matrix} 
\newacronym{HCB}{HCB}{hardcore bosons}
\newacronym{SF}{SF}{spinless fermions}
\newacronym{fRG}{fRG}{functional renormalization group}
\newacronym{LE}{LE}{Luther\hyp Emery}
\newacronym{ASP}{ASP}{adiabatic state preparation}
\newacronym{VQE}{VQE}{variational quantum eigensolver}
\newacronym{METTS}{METTS}{minimally\hyp entangled typical thermal states}
\newacronym{SSH}{SSH}{Su\hyp Schrieffer\hyp Heeger}
\newacronym{GSE}{GSE}{Global Subspace Expansion}
\newacronym{LSE}{LSE}{Local Subspace Expansion}
\newcommand{\ascaddress}{Department of Physics and Arnold Sommerfeld Center for Theoretical Physics (ASC), Ludwig-Maximilians-Universit\"{a}t M\"{u}nchen, D-80333 Munich, Germany}
\newcommand{\mcqstaddress}{Munich Center for Quantum Science and Technology (MCQST), D-80799 M\"{u}nchen, Germany}

\newcommand{\CCQ}{Center for Computational Quantum Physics, Flatiron Institute, 162 5th Avenue, New York, NY 10010, USA}
\newcommand{\saclay}{Universit\'e Paris-Saclay, CNRS, CEA, Institut de Physique Th\'eorique, 91191, Gif-sur-Yvette, France}

\newcommand{\nodagger}[0]{{\vphantom{\dagger}}}
\newcommand{\noprime}[0]{{\vphantom{\prime}}}

\definecolor{corn_flower}{HTML} {78A1E5}
\definecolor{lmugreen}{RGB}{0.0, 148, 64}
\definecolor{Gray}{gray}{0.9}
\newif\ifshowcomments\showcommentstrue

\Crefname{appendix}{Appendix}{Appendices}
\Crefname{equation}{Equation}{Equations}
\Crefname{figure}{Figure}{Figures}
\Crefname{section}{Section}{Sections}
\Crefname{tabular}{Tabular}{Tabulars}
\crefname{appendix}{App.}{Apps.}
\crefname{equation}{Eq.}{Eqs.}
\crefname{figure}{Fig.}{Figs.}
\crefname{section}{Sec.}{Secs.}
\crefname{tabular}{Tab.}{Tabs.}

\definecolor{midnight_blue}{HTML} {002952}
\definecolor{hot_pink}{HTML} {D15D84}
\definecolor{corn_flower}{HTML} {78A1E5}
\definecolor{mauve}{HTML} {905171}
\definecolor{purple}{HTML} {4C3B55}
\definecolor{tiffany_blue}{HTML} {6AA4B0}
\definecolor{gunmetal_grey}{HTML} {4C5355}
\definecolor{honeysuckle}{HTML} {C44B4F}
\definecolor{blueish}{HTML} {226E9C}
\definecolor{bluegray}{HTML} {607D86}
\definecolor{cinnabar}{HTML} {E66354}
\definecolor{walnut}{HTML} {4D181C}
\definecolor{mahagony}{HTML} {4B1816}
\definecolor{aegean_blue}{HTML} {144058}
\definecolor{honey}{HTML} {E58D2E}
\definecolor{persimmon}{HTML} {DD671E}
\definecolor{mimosa}{HTML} {D7A449}
\definecolor{lilac}{HTML} {D5CAE4}

\definecolor{darkteal}{HTML} {3A6D80}
\definecolor{amber}{HTML} {F3CD53}
\definecolor{squash}{HTML} {D56729}
\definecolor{vermillion}{HTML} {9D402D}
\definecolor{cascades_green}{HTML} {355952}
\definecolor{baby_pink}{HTML} {FDC2E4}
\definecolor{wisteria}{HTML} {D3C5E5}
\definecolor{burgundy}{HTML} {800020}

\colorlet{colorUOpO}{midnight_blue}
\colorlet{colorUOpXXV}{persimmon}
\colorlet{colorUOpL}{burgundy}
\colorlet{colorUOpLXXV}{cascades_green}

\definecolor{orcidlogocol}{HTML}{A6CE39}
\tikzset{
  orcidlogo/.pic={
    \fill[orcidlogocol] svg{M256,128c0,70.7-57.3,128-128,128C57.3,256,0,198.7,0,128C0,57.3,57.3,0,128,0C198.7,0,256,57.3,256,128z};
    \fill[white] svg{M86.3,186.2H70.9V79.1h15.4v48.4V186.2z}
                 svg{M108.9,79.1h41.6c39.6,0,57,28.3,57,53.6c0,27.5-21.5,53.6-56.8,53.6h-41.8V79.1z M124.3,172.4h24.5c34.9,0,42.9-26.5,42.9-39.7c0-21.5-13.7-39.7-43.7-39.7h-23.7V172.4z}
                 svg{M88.7,56.8c0,5.5-4.5,10.1-10.1,10.1c-5.6,0-10.1-4.6-10.1-10.1c0-5.6,4.5-10.1,10.1-10.1C84.2,46.7,88.7,51.3,88.7,56.8z};
  }
}

\newcommand\orcidicon[1]{\href{https://orcid.org/#1}{\mbox{\scalerel*{
  \tikzset{external/export=false}
  \begin{tikzpicture}[yscale=-1, transform shape]
    \pic{orcidlogo};
  \end{tikzpicture}
}{|}}}}

\begin{document}

\title{Complex Time Evolution in Tensor Networks}
\author{M.~Grundner\hspace{0.069cm}\orcidicon{0000-0002-7694-0053}}
\email{martin.grundner@physik.uni-muenchen.de}
\affiliation{\ascaddress}
\affiliation{\mcqstaddress}

\author{P.~Westhoff\hspace{0.069cm}\orcidicon{0009-0005-6467-7736}}
\affiliation{\ascaddress}
\affiliation{\mcqstaddress}

\author{F.~B.~Kugler\hspace{0.069cm}\orcidicon{0000-0002-3108-6607}}
\affiliation{\CCQ}

\author{O.~Parcollet\hspace{0.069cm}\orcidicon{0000-0002-0389-2660}}
\affiliation{\CCQ}
\affiliation{\saclay}

\author{U.~Schollw\"ock\hspace{0.069cm}\orcidicon{0000-0002-2538-1802}}
\affiliation{\ascaddress}
\affiliation{\mcqstaddress}

\date{\today}

\begin{abstract}
Real-time calculations in tensor networks are strongly limited in time by entanglement growth, restricting the achievable frequency resolution of Green's functions, spectral functions, self-energies, and other related quantities. By extending the time evolution to contours in the complex plane, entanglement growth is curtailed, enabling numerically efficient high-precision calculations of time-dependent correlators and Green's functions with detailed frequency resolution. Various approaches to time evolution in the complex plane and the required post-processing for extracting the pure real-time and frequency information are compared. We benchmark our results on the examples of the single-impurity Anderson model using matrix-product states and of the three-band Hubbard--Kanamori and Dworin--Narath models using a tree tensor network. Our findings indicate that the proposed methods are also applicable to challenging realistic calculations of materials.  
\end{abstract}

\maketitle

\section{Introduction}
The time evolution of quantum many-body systems such as solids or ultracold gases out of equilibrium is of central interest in physics. Experimental results both in the linear response regime and far from equilibrium may be resolved in real time, such as in pump-probe experiments on solids or in quantum simulations in ultracold atom gases. Time evolution is also reflected in frequency-dependent information, e.g.\ as absorption spectra and inelastic scattering cross-sections, or through the measurement of transport properties. As most of these many-body systems are not amenable to analytical treatment, a large effort has been devoted to developing numerical algorithms for the time evolution of quantum many-body systems. 

In the case of low-dimensional systems, tensor networks \cite{White1992,White1993,Schollwoeck2011,Haegeman2016,Verstraete2008,Paeckel2019} have established themselves as a very powerful descriptive framework. Tensor network algorithms have been proposed which work directly in frequency space \cite{Hall95,Kuhner99,Jeckelmann2002,Weichselbaum2009,Holzner2011,Dargel2012,Wolf2015a,Nocera2016,Baiardi2022} and are typically used for the calculation of spectral functions. A large number of other tensor network methods work directly in real time \cite{Vida03,Vida04,Dale04,Whit04,Vers04a,Zwolak04,Haegeman2011,Zaletel15,Haegeman2016,Paeckel2019}. Frequency-dependent information can then be obtained by Fourier transformations \cite{Whit04}. Most applications have been in the special case of matrix product states. In this paper we will present calculations based both on matrix product states and tree tensor network states. 

The common limitation of all these methods is that they have difficulty in accessing long times or, equivalently, high frequency resolution. This is due to the entanglement barrier: in general, entanglement grows during time evolution. In the worst case, the bipartite von Neumann entanglement entropy $S$ is upper-bounded as \cite{Calabrese2004, Calabrese2005, Eisert2006, Bravyi2007, Osborne2006, Alhambra2021} 
\begin{equation}
    S(t) \leq S(0) + a^{-1} v t ,
\end{equation}
where $v$ is a typical velocity scale of the system, such as the spin wave or Fermi velocity, and $a$ a constant of unit length. As the required resources in tensor networks, typically measured by the bond dimension $m$, scale exponentially with the entanglement, $m \sim e^S$, \cite{Calabrese2004, Calabrese2005} the computational cost of time evolutions then grows exponentially in the worst case \cite{Calabrese2004,Schuch2008} which drastically limits accessible times. 

The worst-case scenario occurs for global quenches of systems. In many cases, however, one is rather interested in numerically more benign time-dependent correlation functions because they relate directly to experimental probes. In the case of zero-temperature calculations for a system with Hamiltonian $\hat{H}$, ground state $\ket{\psi_0}$ and ground state energy $E_0$, they are generically given by
\begin{equation}
    C_{OP}(t) = \bra{\psi_0} \hat{O}^\dagger(t) \hat{P} \ket{\psi_0}
\end{equation}
for two operators $\hat{O}$, $\hat{P}$, where in many applications $\hat{O}=\hat{P}$. The time evolution of  operators is given by
\begin{equation}
    \hat{O}(t) = e^{+i(\hat{H}-E_0)t} \hat{O} e^{-i(\hat{H}-E_0)t} .
\end{equation}
Here, entanglement growth is typically logarithmic in $t$, \cite{Osborne2006,Calabrese2007} leading to a power-law growth of the required resources. While this increases accessible times, these times and hence the achievable frequency resolution are still quite limited. This is a particular problem if one is especially interested in (very) low frequency behavior, which is often the case in solids.

Since their invention, time-evolution methods for tensor networks have also been frequently used on the imaginary axis, for instance for ground-state searches. This reflects the fact that imaginary time evolution generally acts as a gradual energy truncation: it suppresses high-energy states and enhances low-energy states exponentially in the long-time limit \cite{Haegeman2016}. Entanglement growth does not act as a limiting factor.  So far, energy truncation algorithms have only been combined successfully with tensor network methods that calculated correlation function directly in frequency space \cite{Holzner2011,Wolf2015a}. Imaginary-time evolution methods for tensor networks have also been successfully employed in the context of dynamical mean-field theory (DMFT) and related embedding methods \cite{Georges1992,Georges1996,Kotliar2006} giving access to realistic multiband models, but with limited reliable information on low-frequency behavior \cite{Wolf2015b,Linden2020,Bramberger2021,Bramberger2023}. Related real-time methods for DMFT have made good progress over the last years \cite{Ganahl2014,WolfParcollet2014,Ganahl2015,Bauernfeind2017} but realistic materials with multiband physics are very challenging and frequency resolution is limited by finite simulation times due to entanglement growth.    

The goal of this paper is to combine real and imaginary time evolutions in the complex plane in order to develop methods to extract real-time information while using imaginary-time evolution to limit the growth of entanglement during time evolution. This leads to much faster, less resource-intensive calculations which in turn give cheaper access to the same information as real-time evolutions or allow one to proceed to longer times and better frequency resolution. Note that the idea of complex time evolution has been recently discussed in the context of Quantum Monte Carlo simulations \cite{Guther2018}. Similar ideas are also published in another work \cite{Cao2023}.  

We introduce contours in the complex plane as $z(t)=t+i\tau(t)$ and extend the definition of $C_{OP}(t)$ as 
\begin{equation}
    C_{OP}(z) = \bra{\psi_0}\hat{O}^\dagger (z) \hat{P}\ket{\psi_0},
\end{equation}
where
\begin{equation}\label{def:ComplexTimeEvolution}
\hat{O}(z) = e^{+i(\hat{H}-E_0)z} \hat{O} e^{-i(\hat{H}-E_0)\overline{z}}
\end{equation}
with $\overline{z}$ the complex-conjugate of $z$. Effectively, then 
\begin{equation}
    C_{OP}(z) = \bra{\psi_0}\hat{O}^\dagger e^{-i(\hat{H}-E_0)\overline{z}} \hat{P}\ket{\psi_0}.
\end{equation}
Any discussion of possible contours in the complex plane cannot be exhaustive;
we present easily implemented contours for which reliable post-processing
methods exist to extract the desired real-time result. The various schemes 
allow mutual verification, and present a different balance of speed vs precision.
We first focus on the single-impurity
Anderson model (SIAM) as a test case for which extremely precise data is
available, also within the framework of tensor network methods. After
establishing benchmark data (Sec.~\ref{sec:benchmarkSIAM}), we introduce two
main classes of contours: a contour
parallel to the real axis
(Sec.~\ref{sec:parallel}) and a tilted contour inclined with respect to the real
axis (Sec.~\ref{sec:inclined}). The results from these contours all have to be post-processed, for which we present various approaches; we will denote the resulting methods by ``contour used (post-processing approach)''. We show results both
for the impurity spectral function and for the numerically challenging
calculation of the self-energy at very low frequencies, down to the Fermi
liquid regime where $\text{Im}\,\Sigma(\omega) \sim \omega^2$. An additional
procedure is based on a kink contour; it requires no post-processing at all,
but has to be controlled for numerical instability; it is discussed in
Appendix~\ref{real_w_ac}, but not used in the main text. All methods exhibit a
substantial speed-up and provide accurate results. The methods are robust and
cheap. Furthermore, unlike in \cite{Cao2023}, our methods do not require to calculate
powers of the Hamiltonian, a task which may prove difficult for multiband systems.
We further illustrate our method on the
example of a model with three impurity orbitals that interact via a
Hubbard--Kanamori or a Dworin--Narath interaction (Sec.~\ref{sec:threeband}). We show that even the
calculation of the self-energy at very low frequencies can be achieved in the
three-band case, with a quality comparable to the Numerical Renormalization Group (NRG) \cite{Wilson1975,Krishna1980,Bulla2008}.
We therefore expect that our methods are also applicable in realistic calculations for multiband materials. 

\section{Benchmark model: SIAM}\label{sec:benchmarkSIAM}
The first test case for the usefulness of the various complex time-evolution schemes is the retarded impurity Green's function and the associated impurity spectral function of the single impurity Anderson model (SIAM). The SIAM is a prototypical model describing magnetic impurities in metals \cite{Anderson1961} that features low energy structures that are well understood both numerically \cite{Wilson1975,Krishna1980,Bulla2008,Raas2005} and analytically \cite{Anderson1961,Schrieffer1966}. 

The Hamiltonian of the SIAM is given by $\hat{H}=\hat{H}_{\mathrm{imp}}+ \hat{H}_{\mathrm{bath}}$ with
\begin{align}	\hat{H}_{\mathrm{imp}}=&U\hat{n}^{\nodagger}_{0\uparrow}\hat{n}^{\nodagger}_{0\downarrow} + \epsilon^{\nodagger}_{0}(\hat{n}^{\nodagger}_{0\uparrow}+\hat{n}^{\nodagger}_{0\downarrow}), \\
   \hat{H}_{\mathrm{bath}}=& \sum_{k=1}^{N_{\mathrm{b}}} \sum_{\sigma}\big(v^{\phantom{ast}}_{0,k\sigma}\hat{c}^{\dagger}_{k\sigma}\hat{c}^{\nodagger}_{0\sigma}+v^{\ast}_{0,k\sigma}\hat{c}^{\dagger}_{0\sigma}\hat{c}^{\nodagger}_{k\sigma}\big) \nonumber \\ &+ \sum_{k=1}^{N_{\mathrm{b}}} \sum_{\sigma}\epsilon^{\nodagger}_{k\sigma}\hat{n}{\nodagger}_{k\sigma},
\label{eq:SIAM}
\end{align}
where $\hat{c}^{\dagger}_{k\sigma}$ ($\hat{c}^{\nodagger}_{k\sigma}$) are fermionic creation (annihilation) operators with spin $\sigma \in \{\uparrow, \downarrow\}$. The impurity site is located at $k=0$. $\hat{n}^{\nodagger}_{k\sigma}$ are density operators and $U$ is the Coulomb repulsion strength. $\epsilon_0$ is chosen as $-U/2$; the model is at half\hyp filling. The on\hyp site potentials $\epsilon^{\nodagger}_{k\sigma}$ and the impurity-bath hopping elements $v^{\phantom{ast}}_{0,k\sigma}$ are obtained by discretizing the hybridization function $\Delta(\omega)$. For our calculations we use a semi\hyp elliptical hybridization function defined as 
\begin{equation}
	-\frac{1}{\pi}\text{Im}\Delta(\omega)= \frac{D}{2\pi}\sqrt{1-\bigg(\frac{\omega}{D}\bigg)^{2}},
\end{equation}
for $\omega\in[-D,D]$, where $D$ represents the half\hyp band width. The support of the hybridization function is discretized linearly into an odd number of equally sized intervals; see  \cite{Bulla2008,Ines2015} for details of this discretization procedure. For simplicity, we will drop all spin indices in the following.  

The $T=0$ retarded impurity Green's function is, without spin indices, given by 
\begin{equation}
    G(t)=-i\theta(t)\expval{\{\hat{c}^{\nodagger}_{0}(t),\hat{c}^{\dagger}_{0}\}}{\psi_0} ,
\end{equation}
where $\ket{\psi_0}$ is the ground state of the SIAM with energy $E_0$. A Fourier transformation yields
\begin{equation}
    G(\omega) = \int_0^\infty \, e^{i\omega t}e^{-\eta t} G(t) ,
\end{equation}
where $\eta$ is the usual damping factor. By definition, the impurity spectral function is then given by
\begin{equation}
\mathcal{A}(\omega) = -\frac{1}{\pi}\textrm{Im}\, G(\omega) .
\end{equation}

The calculation of $G(t)$ requires a real-time evolution.
For time-independent Hamiltonians as here the problem of entanglement growth in the calculation of time-dependent correlators by matrix product state methods can be reduced by exploiting the homogeneity in time \cite{Karrasch2013,Barthel2013}, e.g.\
\begin{equation}
    \bra{\psi_0} \hat{c}^\dagger (0) \hat{c}^\nodagger(t) \ket{\psi_0} =
    \bra{\psi_0} \hat{c}^\dagger (-t') \hat{c}^\nodagger(t'') \ket{\psi_0}, 
\end{equation}
where we have split $t>0$ into $t=t'+t''$ with $t', t''>0$. This can be written as
\begin{equation}
    \bra{\psi_0} \hat{c}^\dagger (0) \hat{c}^\nodagger(t) \ket{\psi_0} = 
    [\hat{c}^\nodagger(-t') \ket{\psi_0}]^\dagger \hat{c}(t'') \ket{\psi_0}.
\end{equation}
The real-time evolution up to $t$ has been split into two time evolutions up to $-t'$ and $t''$, respectively, and a final overlap which also occurs in the original formulation. Optimal limitation of entanglement growth is typically achieved for a symmetric split $t'=t''=t/2$. The following real-time calculations all use this symmetric split. 

For the time evolution we used the two\hyp site version of the Time\hyp Dependent Variational Principle (2TDVP) \cite{Haegeman2011,Haegeman2016,Paeckel2019}.
All results were obtained using a time step of $\delta t = 0.2 D^{-1}$ and a truncated weight (the sum of the discarded reduced density matrix eigenvalues) of $w_{t}=10^{-11}$, which we checked to give a converged result; the MPS bond dimension is allowed to grow accordingly. Up to the maximum time of $t_{\textrm{max}}=90 D^{-1}$ the calculations reached a maximum bond dimension of $m\sim 1500$ for $w_t=10^{-10}$ and $m\sim 2700$ for $w_t=10^{-11}$. We chose values $U/D=2$ and as bath size $N_b=59$, except when stated otherwise. In the Fourier transformation, we used $\eta =0.001 D$ for spectral functions, except if stated otherwise, and $\eta=0$ for self\hyp energies. All tensor network simulations were performed using the \textsc{SyTen} toolkit \cite{Syten}.

\section{Contour at constant imaginary time: Parallel contour}\label{sec:parallel}

\begin{figure}
  \centering
  \includegraphics[width=0.9\linewidth]{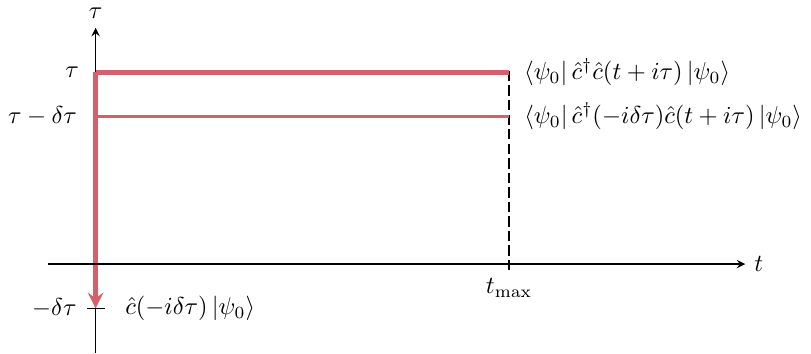}
  \caption{
    \label{fig:contours1}
    Parallel contour $t + i\tau$.
Multiple complex parallel contours are obtained with a small shift $\delta \tau$ from a first contour, cf.\ text.
}
\end{figure}

We now turn to the first complex-time method, where time is evolved along a
complex contour parallel to the real-time axis, i.e. at constant imaginary
time, see Fig.~\ref{fig:contours1}. We have $z(t)=t+i\tau$ with
$t\in[0,t_{\mathrm{max}}]$ and $\tau>0$ constant. Here, $z$ starts at
$0$, moves up to $i\tau$, and then continues from $i\tau$ to
$t_{\textrm{max}}+i\tau$.
The entanglement growth with time is strongly limited compared the real time evolution, 
as illustrated in \cref{comp_contours} in Appendix~\ref{app:comp_contours},
where we compare the time dependency of the entanglement for all the contours introduced in this paper.
This contour suppresses high energy states aggressively, as reflected in the 
small entanglement growth at short times compared to other contours.
It is therefore advantageous to use comparatively small $\tau>0$
to retain a sufficient amount of high energy spectral weight. 

An attractive feature of the parallel contour is that additional parallel contours offset by $\delta \tau$ can be obtained very cheaply from a first contour.
Let us illustrate this point with one correlator
\begin{multline}
    \bra{\psi_0} \hat{c}^\dagger \hat{c}^\nodagger(t+i(\tau+\delta\tau)) \ket{\psi_0} \nonumber \\ 
   = [e^{-(\hat{H}-E_0)\delta \tau}\hat{c}\ket{\psi_0}]^\dagger e^{i(\hat{H}-E_0)(t+i\tau)} \hat{c} \ket{\psi_0} .
\end{multline}
We first compute the time evolution of the right ket up to time $t + i\tau$, and 
the additional time evolution $\delta \tau$ for the left bra, which is cheap to compute for small $\delta \tau$.
Note that we can use positive or negative $\delta \tau$. In practice, we use 
$\delta \tau< 0$, as illustrated on Fig.~\ref{fig:contours1}. This allows us to obtain the correlations 
on a contour closer to the real axis, which is a priori harder to compute, at the price of a positive exponential 
evolution of the left term for a short time $\delta \tau$.

Let us introduce the complex-time spectral function $\mathcal{A}_\tau$
defined by 
\begin{equation}\label{defAtau}
\hspace{-0.2cm}
    \mathcal{A}_\tau(\omega) \equiv \frac{1}{2\pi} \int dt e^{i (\omega+i\eta) t} \expval{\{\hat{c}^{\nodagger}_{0}(t + i\tau),\hat{c}^{\dagger}_{0}\}}{\psi_0}.
\end{equation}

At $\tau=0$, this function is the ordinary spectral function 
$ \mathcal{A}_{\tau=0}(\omega) = \mathcal{A}(\omega) =  -\frac{1}{\pi}\,\text{Im} \, G^R(\omega)$.
Due to the simple form of the contour, we have an explicit relation between 
$ \mathcal{A}_\tau$ and $\mathcal{A}$:
\begin{equation}\label{inversion_eq}
    \mathcal{A}_{\tau}(\omega) = \mathcal{A}(\omega)e^{-\tau|\omega|}
\end{equation}
which is established in Appendix \ref{app:AtauRelationProof} and exact in the limit $\eta\rightarrow0$ as is used here. Our method consists simply in computing $\mathcal{A}_\tau(t)$,
then $\mathcal{A}_\tau(\omega)$ using a Fourier transform, 
and finally invert \eqref{inversion_eq} to obtain the spectral function $\mathcal{A}(\omega)$.
Despite its apparent simplicity, this inversion presents however
two difficulties, both at low and high frequencies.

First, the inversion of \eqref{inversion_eq} is clearly difficult at high frequencies, 
as small errors in $\mathcal{A}_{\tau}(\omega)$ are amplified by the exponential.
Such errors may result from a too small $t_{\mathrm{max}}$ for $\mathcal{A}_{\tau}(t)$ to be fully decayed, 
or from a broadening in the Fourier transform.
This issue can be solved by introducing a cut\hyp off frequency $\omega_{c}$ that limits the growth of the exponential factor as \
\begin{equation}
    \mathcal{A}(\omega) = \mathcal{A}_{\tau}(\omega)e^{\tau\min(|\omega|,\omega_c)},
\end{equation}
where $\omega_{c}$ is chosen such that it only acts in the high frequency tail of the spectral function. We found $\omega_c=3D$ to yield overall good results in our calculations.

Second, the inversion of \eqref{inversion_eq} is also difficult at low frequencies, 
even though the exponential term is close to 1.
In this work, and in many physics applications, we are interested in a high precision computation of the 
behavior of the spectral function $\mathcal{A}(\omega)$ or the self-energy $\Sigma(\omega)$ at low frequencies.
The difficulty comes from the kink in  $\mathcal{A}_{\tau}(\omega)$ at $\omega = 0$ in Eq. (\ref{inversion_eq}), 
caused by $|\omega|$ in $e^{-\tau|\omega|}$.
This is an exact feature, but as $\mathcal{A}(\omega)$ does not exhibit a kink
at $\omega=0$, this must be exactly compensated by a kink in $\mathcal{A}_\tau(\omega)$
at $\omega=0$. The latter requires high-accuracy results from long-time
evolutions where tensor network methods are limited. As a result, the compensation is
imperfect. 

Our solution of this issue uses a linear combination of a few $\mathcal{A}_{\tau_k}(\omega)$
computed on  parallel contours $t + i\tau_k$ (as described above), which is designed to compensate the low frequency singularity
introduced by the $e^{-\tau|\omega|}$ factor.
Introducing some weights $a_k$, and defining the function
\begin{equation}
   h(\omega) \equiv \sum_{k=1}^n a_k e^{-\tau_k|\omega|} 
\end{equation}
we have
\begin{equation}\label{eq:A_ak}
    \mathcal{A}(\omega) h(\omega) = \sum_{k=1}^n a_k \mathcal{A}_{\tau_k}(\omega).    
\end{equation}
We choose the weights $a_k$ such that the function $h(\omega)$ is flat close to $\omega=0$, 
i.e. we cancel the first powers of its low $\omega$ expansion:
\begin{subequations}\label{eq:coef_ak}
\begin{align}
   \sum_{k=1}^n a_k &= 1
   \\
   \sum_{k=1}^n a_k \tau_k^l &= 0
\end{align}
\end{subequations}
for $l=1, \ldots, n-1$. 
As $h$ is close to 1 at low $\omega$, $h(\omega) = 1 + O(\omega^{n})$,  we can now safely invert \eqref{eq:A_ak}:
\begin{equation}
    \mathcal{A}(\omega) =  h^{-1}(\omega)\, \sum_{k=1}^n a_k  \mathcal{A}_{\tau_k}(\omega). 
\end{equation}
In practice, we need to take $\tau_k$ which are neither too close 
(leading to large $a_i$ that amplify numerical noise, due to the Vandermonde determinant in the linear system
\eqref{eq:coef_ak}), nor too distant as time-evolution errors affect
precision. 
In this work, we used $n=3$ and $\tau_k=1, 1.15, 1.3$, if not stated otherwise. We will refer to this approach as the parallel (inversion) method.

Note that one can also perform a direct extrapolation to $\tau =0$ using several parallel contours, which we call the parallel (extrapolation) method.
We present it in Appendix~\ref{app:extrapolationmethod} for completeness, but
it turns out to be slightly inferior to the parallel (inversion) method.
The parallel (extrapolation) method will be used in this paper as a check for the parallel (inversion) method.


\begin{figure}
  \centering
  \subfloat[\label{fig_inversion}]{
    \includegraphics[width=0.8\linewidth]{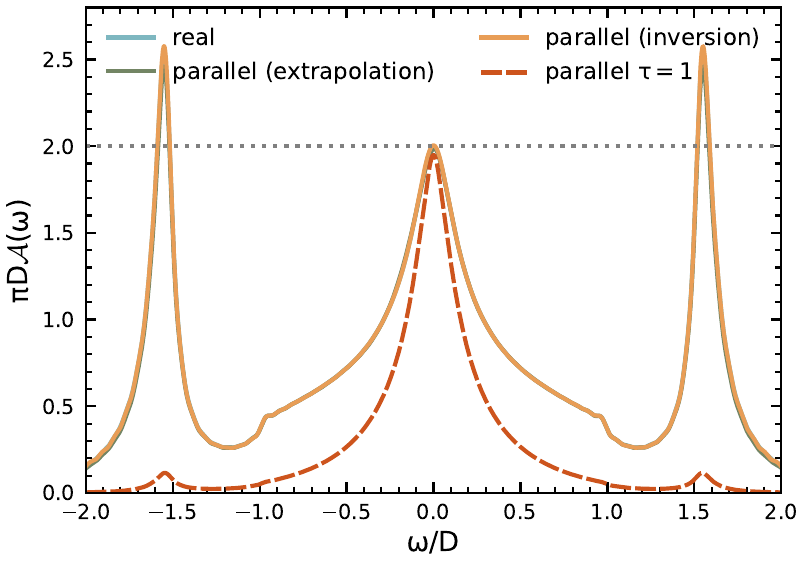}
  }%
  \\
  \subfloat[\label{parallel:zoom}]{
    \includegraphics[width=0.8\linewidth]{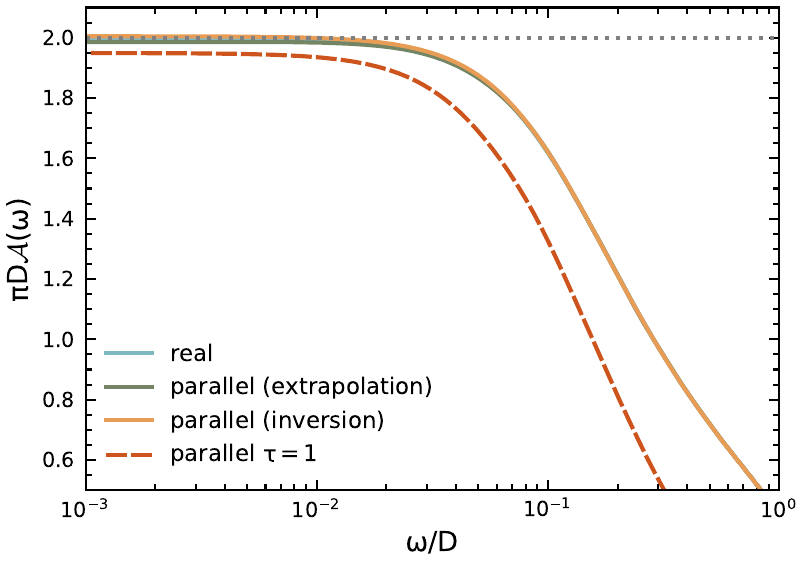}
  }
  \caption{
    \label{fig:inversion_full}
    \subfigref{fig_inversion} Comparison of spectral functions obtained on a
    parallel contour: "real" stands for the real-time reference data, 
    $\tau=1$ on the parallel contour without post-processing, parallel (extrapolation) and parallel (inversion) for the two methods of the main text. 
    Nearly all spectral weight can be recovered by both methods in excellent agreement. 
    The parallel (inversion) method used $n=3$ with $\tau_k=1, 1.15, 1.3$; the coefficients $a_k$ (see text) are then approximately $33.2, -57.8, 25.6$.
    The parallel (extrapolation) method used corrections to 6th order with 13 contours
    centered on $\tau=1$ with a distance $\delta \tau=0.075$.
    \subfigref{parallel:zoom} Zoom at low frequencies with the same labels, 
    showing the high accuracy of the parallel (inversion) and parallel (extrapolation) methods at very low frequencies.
    }
\end{figure}

\begin{figure}
  \centering
  \subfloat[\label{fig:inversion_siam_specs}]{
    \includegraphics[width=0.8\linewidth]{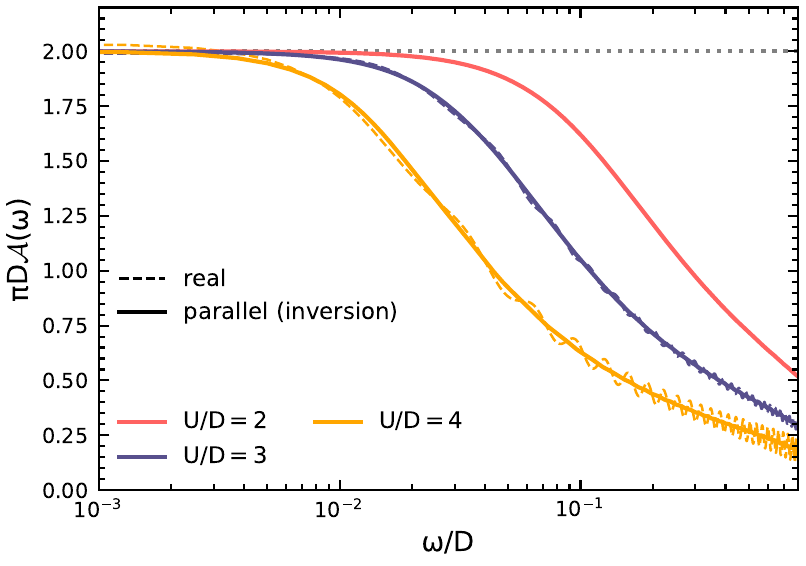}
  }%
  \\
  \subfloat[\label{fig:inversion_siam_selfe}]{
    \includegraphics[width=0.8\linewidth]{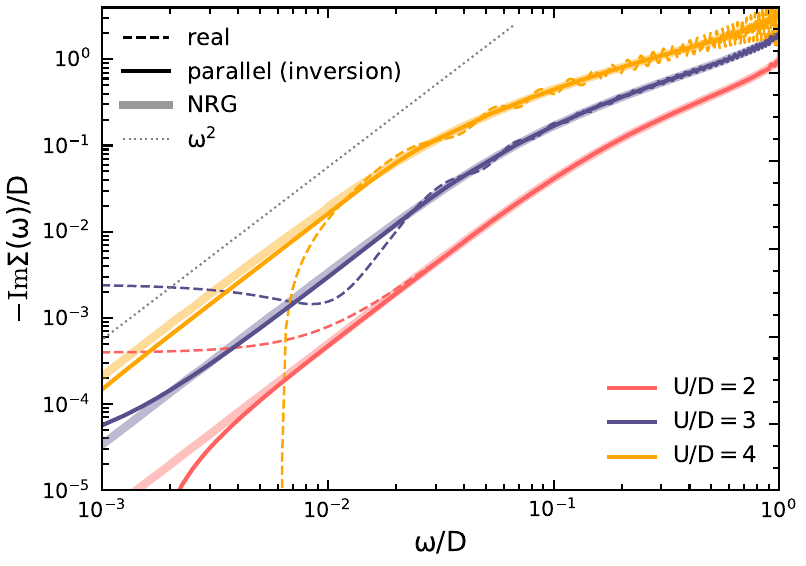}
  }
  \caption{
    \label{fig:Inversion_Us}
    \subfigref{fig:inversion_siam_specs} Spectral function of the SIAM for
    $U/D\in\{2,3,4\}$ with $N_b=299$, using the $n=2$ parallel (inversion) method with
    $\tau_k\in\{1.15, 1.3\}$; the coefficients are approximately $a_k\in\{8.6,
    -7.6\}$. We have an overall higher frequency resolution due to a finer bath
    discretization, zero broadening and a time evolution
    $t_{\textrm{max}}\in\{150D^{-1},200D^{-1},250D^{-1}\}$. 
    The maximum bond dimension is $m=1500$ $SU(2)$ states. In all cases, in particular for larger $U/D$, the matching of
    the Friedel sum rule is less accurate for the real-time method.
    \subfigref{fig:inversion_siam_selfe} Corresponding self\hyp energies. 
 }
\end{figure}

{\em Results.} 
We present results of the spectral function of the benchmark SIAM in Figs.~\ref{fig_inversion} and \ref{parallel:zoom}.
We use $\delta t=0.2D^{-1}$, $w_t=10^{-11}$, $N_b=59$, $t_\text{max}=90D^{-1}$, and $\eta=0.001 D$. The simulation reached a maximum bond dimension of $m=1023$. 
Both the parallel (inversion) and parallel (extrapolation) method provide spectral functions in
excellent agreement with the much more costly (more than an order of magnitude)
real-time benchmark data both at higher and in particular also at very low
frequencies. The (generalized) Friedel sum rule $\pi D \mathcal{A}(0)=2$ is matched to fractions of a percent. 
As both methods operate on the same numerical data, the parallel (extrapolation) method
provides a cheap control of the quality of the spectral function. For Fig.~\ref{fig:inversion_siam_specs} we use larger baths and a larger maximum bond dimension to calculate the spectral function for larger values of $U/D$ both by real-time calculations and by the parallel (inversion) method. The Friedel sum rule is matched to very high accuracy in all cases; real-time results are much less accurate for larger $U/D$.

The determination of the self-energy $\Sigma(\omega)$ is numerically more challenging than that of the spectral function $\mathcal{A}(\omega)$. We calculate it 
from the Dyson equation $\Sigma(\omega) = G_0^{-1}(\omega)-G^{-1}(\omega)$, where $G_0(\omega)$ is the impurity Green's function for the non-interacting case. Its inverse is given by 
$G_0(\omega)^{-1}=\omega+i\eta -\epsilon_0 -\Delta(\omega)$ with $\eta=0$ here. We used the analytically known hybridization function. We obtain $G(\omega)$ from the parallel (inversion) method, which yields $\mathcal{A}(\omega)$, i.e.\ the imaginary part of $G(\omega)$. We then use the Kramers--Kronig rule to obtain the real part of $G(\omega)$.

To assess the quality of our self-energies, we henceforth include benchmark results from NRG.
The NRG data was obtained in a state-of-the-art implementation
\cite{Weichselbaum2007,Peters2006,Zitko2009,Weichselbaum2012,Lee2017,Lee2016}
based on the QSpace tensor library \cite{WeichselbaumAnnnals2012}, using a
symmetric improved estimator for the self-energy \cite{Kugler2022}.
The latter allows one to follow the imaginary part of the self-energy down to extremely low values of $|\Im\Sigma|/D$.

As shown in Fig.~\ref{fig:inversion_siam_selfe}, the self energies calculated
by complex time evolution reach the Fermi-liquid $\omega^2$-regime. The final
breakdown at very low frequencies ($\omega/D\approx 0.007$) is due to small
deviations from the sum rule and the use of Dyson's equation. We expect that
the use of improved estimators \cite{Kugler2022} will alleviate this problem in
future implementations. Before the breakdown, agreement with NRG results is
excellent. Note that self-energies calculated from real-time results are much
less accurate and plagued by unphysical oscillations.


\section{Complex-time evolution at fixed angle: tilted contour}\label{sec:inclined}

\begin{figure}[h]
    \centering
    \includegraphics[width=0.6\linewidth]{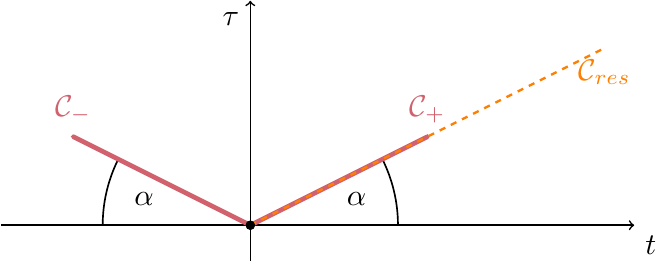}
    \caption{Symmetrically split complex-time contour: Contour $\mathcal{C}_{\mathrm{res}}$ from 0 to $z$ is replaced by two symmetric contours $\mathcal{C}_+$ and $\mathcal{C}_-$ from $0$ to $z/2$ and $-\overline{z}/2$ respectively.
    } 
    \label{fig:splitsimplecomplexcontour}
\end{figure}

In our second complex-time approach, time is evolved along a complex contour tilted by various angles $\alpha$ with respect to the real axis, where
$z(t)= t + i t \tan \alpha$, i.e. $\tau=t \tan \alpha>0$. We adapt the symmetric splitting of real\hyp time evolutions to the complex plane by splitting $z=z'+z''$ with $z'=z''=z/2$.   
Then we have 
\begin{align}\label{complex_time_split}
\bra{\psi_0} \hat{c}^\dagger (0) \hat{c}^\nodagger(z) \ket{\psi_0} &=  [\hat{c}^\nodagger(-\overline{z}/2) \ket{\psi_0}]^\dagger \hat{c}^\nodagger(z/2) \ket{\psi_0}, \\
\bra{\psi_0} \hat{c}^\nodagger (z) \hat{c}^\dagger(0) \ket{\psi_0} &=  [\hat{c}^\dagger(z/2) \ket{\psi_0}]^\dagger \hat{c}^\dagger(-\overline{z}/2) \ket{\psi_0}.
\end{align}
The two contours on which complex time evolution occurs now look as in Fig.~\ref{fig:splitsimplecomplexcontour}. 

\begin{figure}
  \centering
  \subfloat[\label{fig:comp_alphas}]{
    \includegraphics[width=0.8\linewidth]{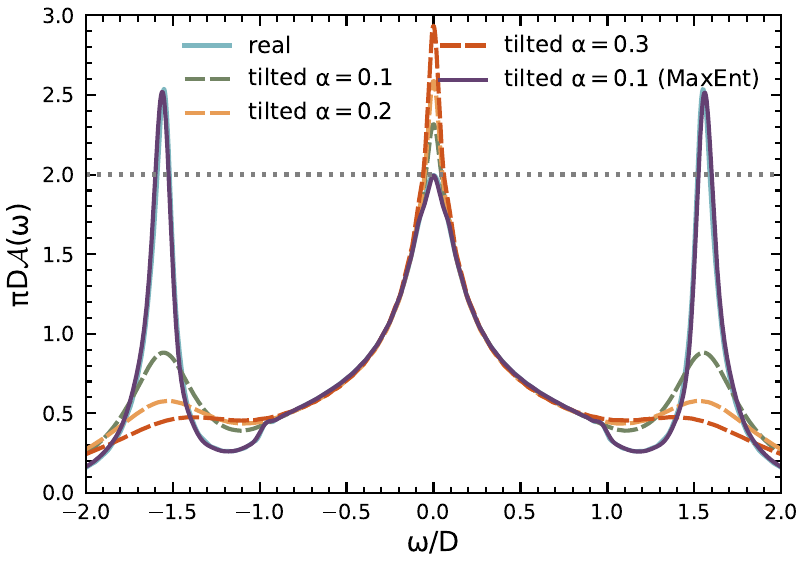}
  }%
  \\
  \subfloat[\label{fig:comp_alphas_log}]{
    \includegraphics[width=0.8\linewidth]{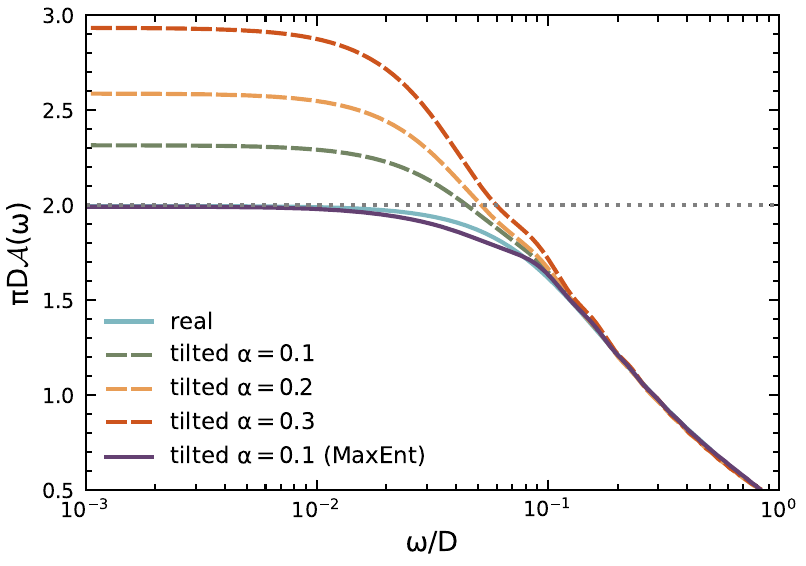}
  }
  \caption{
    \label{fig:comparison_alphas_full}
    \subfigref{fig:comp_alphas} Comparison of spectral functions obtained on tilted contours with real-axis reference data ('real') 
    and the tilted (MaxEnt) result from the tilted contour with $\alpha=0.1$. The kink at $\omega/D=\pm 1$ is model specific. The dashed lines are data before post\hyp processing.
    \subfigref{fig:comp_alphas_log} shows the spectral functions of
    \subfigref{fig:comp_alphas} and $\mathcal{A}(\omega)$ obtained after a
    MaxEnt continuation of $G_{0.1}(z)$ on a logarithmic frequency scale for
    $\omega/D>0$ in $[10^{-3},10^{0}]$. 
    }
\end{figure}

\begin{figure}
    \centering
    \includegraphics[width=0.9\linewidth]{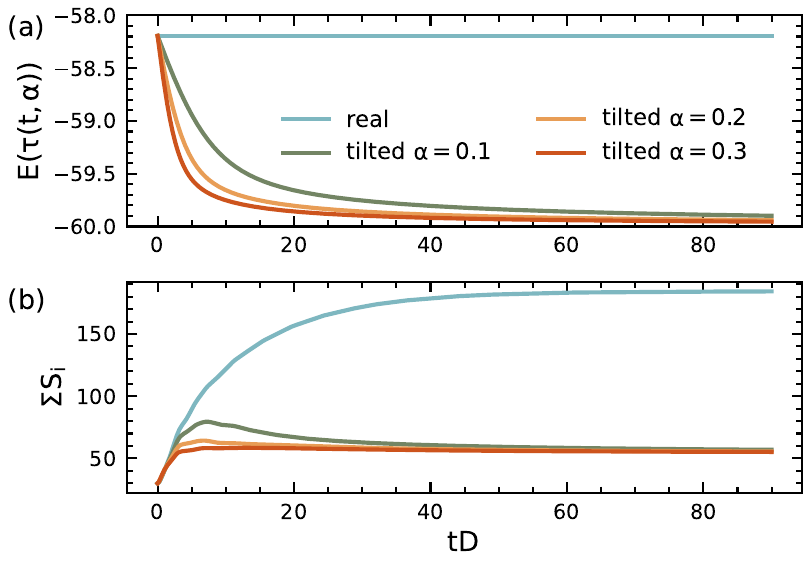}
  \caption{
    \label{fig:comp_energy_shannon}
    Energy (a) and entanglement entropy (b) during time-evolution for $U/D=2$. Imaginary time evolution along a tilted contour shifts the energy expectation value of the time-evolving state towards the ground state energy. Entanglement entropy, here given by the sum of the entanglement entropies for all system cuts,  is reduced strongly along the tilted contours resulting in substantial speed\hyp ups.
    }
\end{figure}

In all calculations, we use $\delta |z| =0.2 D^{-1}$, and again $w_t = 10^{-11}$ and $N_b=59$. The maximum complex time is in all cases $|z|_{\mathrm{max}}=90 D^{-1}$, or real times $t_{\mathrm{max}}=90 D^{-1} \cos \alpha$. The resulting Green's function is Fourier transformed as
\begin{equation}
    G_\alpha(\omega) = \int_{0}^\infty dt \, G_\alpha(t+i\tau) e^{i\omega t - \eta t}
\end{equation}
(again with $\eta=0.001D$); the subscript $\alpha$ indicates that we evaluate on a tilted contour. The spectral function $\mathcal{A}(\omega)$ is then
extracted as $\mathcal{A}_\alpha(\omega)\equiv -(1/\pi)\text{Im} G_\alpha(\omega)$. Results
are shown on a linear and a logarithmic frequency scale in
Fig.~\ref{fig:comparison_alphas_full}. Note that we do not expect to obtain the
correct $\mathcal{A}(\omega)$ for $\alpha \neq 0$, as $\mathcal{A}_\alpha$ is just an intermediate step in the computation.
Nevertheless, we find that peak positions in $\mathcal{A}_\alpha$ are well preserved for moderate
angles thus still allowing for a rough interpretation of quasi-particle peaks,
see Fig.~\ref{fig:comp_alphas}.
We find a significant reduction in the
entanglement entropy for all angles that can be understood as the result of
damping the statistical weight of high energy states, as it is reflected in a
reduction in the time\hyp dependent energy expectation value of the system, see
Fig.~\ref{fig:comp_energy_shannon}. The dampened growth in the entanglement
entropy leads to a speed\hyp up of a factor of the order 100 in the SIAM model
for the specified parameters compared to our real\hyp time reference
calculation. This is reflected in the very small final bond dimensions, $m=52$
for $\alpha=0.3$, $m=66$ for $\alpha=0.2$ and $m=137$ for $\alpha=0.1$.

In order to compute the spectral function $\mathcal{A}(\omega)$, the data has to be post-processed, in this case by an analytical continuation of the complex-time data to the real axis by 
MaxEnt, yielding the tilted (MaxEnt) method; see \cref{ac_sec} for a detailed discussion of MaxEnt. The
agreement with the real-time data is overall very good and the Friedel sum rule
obeyed to $0.4\%$, but there are small, but noticeable deviations from the true
$\mathcal{A}(\omega)$ between $\omega/D\sim 10^{-2}$ and $\omega/D\sim 10^{-1}$
(Fig.~\ref{fig:comp_alphas_log}). As the analytic continuation kernel is
generally ill-conditioned and its current formulation leads to deviations
around the Fermi edge, this is not surprising. In comparison to the parallel (inversion) and
parallel (extrapolation) methods, accuracy is lower for the tilted (MaxEnt) method, in particular
at low frequencies. The tilted (MaxEnt) method is however the fastest, 
so may be very useful in DMFT applications for the intermediate steps of the iteration procedure. Note that the tilted contour can also be combined with the extrapolation scheme of Appendix~\ref{app:extrapolationmethod} to yield the tilted (extrapolation) method; results will be shown for the self-energies calculated in the following section.

\section{Three-band model calculations}
\label{sec:threeband}
The need for reliable high-performance calculations of low-frequency
information is particularly pressing for multiorbital models used
in realistic quantum embedding computations of strongly correlated materials \cite{Georges1992,Georges1996,Kotliar2006}, 
more specifically for transport computations.
Accurate real frequencies quantum impurity solvers are rare, with the notable exception of NRG, 
which is however limited in the number of orbitals.
In order to benchmark our method in such a case, 
we use the three-band Anderson model with the Hubbard--Kanamori interaction, which reads
\begin{align}
&\hat{H}_{K} =\; U\sum_{m}\hat{n}_{m^{\noprime}\uparrow}\hat{n}_{m^{\noprime}\downarrow}+U^{\prime}\sum_{m^{\noprime}\neq m^{\prime}}\hat{n}_{m^{\noprime}\uparrow}\hat{n}_{m^{\prime}\downarrow}+ \nonumber \\ 
&+(U^{\prime}-J)\sum_{m^{\noprime}<m^{\prime},\sigma}\hat{n}_{m^{\noprime}\sigma}\hat{n}_{m^{\prime}\sigma} - J\sum_{m^{\noprime}\neq m^{\prime}}\hat{d}^{\dagger}_{m^{\noprime}\uparrow}\hat{d}^{\nodagger}_{m^{\noprime}\downarrow}\hat{d}^{\dagger}_{m^{\prime}\downarrow}\hat{d}^{\nodagger}_{m^{\prime}\uparrow}  \nonumber\\
&+J\sum_{m\neq m^{\prime}}\hat{d}^{\dagger}_{m^{\noprime}\uparrow}\hat{d}^{\dagger}_{m^{\noprime}\downarrow}\hat{d}^{\nodagger}_{m^{\prime}\downarrow}\hat{d}^{\nodagger}_{m^{\prime}\uparrow} 
\label{eq:Hubbard-Kanamori}
\end{align}
where $m, m'$ run from 1 to 3, $U$ and $U^{\prime}$ are the intra- and
inter\hyp orbital Hubbard interaction, and $J$ is the Hund's coupling.
$\hat{d}_{m\sigma}$ and $\hat{d}^\dagger_{m\sigma}$ are fermionic annihilation
and creation operators on band $m$. For the sake of consistency with real
material calculations, we choose $U^{\prime}=U-2J$ \cite{Georges2013}.
Each impurity couples to a bath as in the SIAM. 
While it is not the most generic case (which would have  non-diagonal bath couplings), 
we expect our approach to generalize without major difficulty.
\begin{figure}[h]
    \centering
    \includegraphics[width=0.9\linewidth]{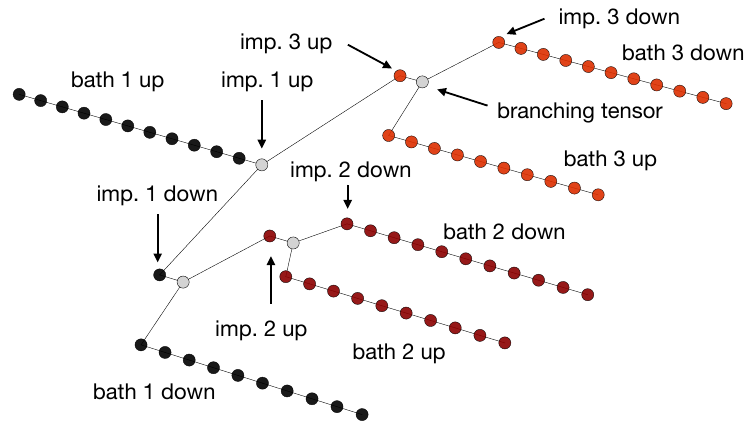}
    \caption{T3NS for three-orbital calculations: The physical basis of all 3 impurities and their associated bath sites are split into spin-up and spin-down representations (empty or occupied). The resulting 6 (spinless) impurity orbitals are connected by branching tensors. The bath sites are attached as MPS-like chains.}
    \label{fig:T3N_sketch}
\end{figure}

As density of states we use the same semi-circular density of states as before and obtain the bath parameters as for the SIAM. Each bath is modeled by $N_b=99$ bath sites, if not stated otherwise. As interaction parameters we consider $U/D=2$ and $J/D=0.3$. In all calculations, we set $\eta=0$.

The simulations are carried out using 2TDVP until $t=20D^{-1}$ and 1TDVP until the final time of $t=180D^{-1}$. We use a tree tensor network, more specifically a T3NS \cite{Gunst18} where trees are constructed only of rank-three tensors. These are either branching tensors with three legs establishing a tree geometry or physical tensors identical to matrix-product state tensors with two auxiliary connecting legs and one physical leg carrying the physical degrees of freedom (see Fig.~\ref{fig:T3N_sketch}). 
This representation is more adequate to multiorbital impurity problems and contains the fork geometry of \cite {Bauernfeind2017} as a special case without the need of numerically costly four-leg tensors. The TDVP implementation follows \cite{Bauernfeind2020}. 

\begin{figure}[H]
  \centering
  \subfloat[\label{fig:HK_global}]{
    \hspace*{-0.2em}
    \includegraphics[width=0.9\linewidth]{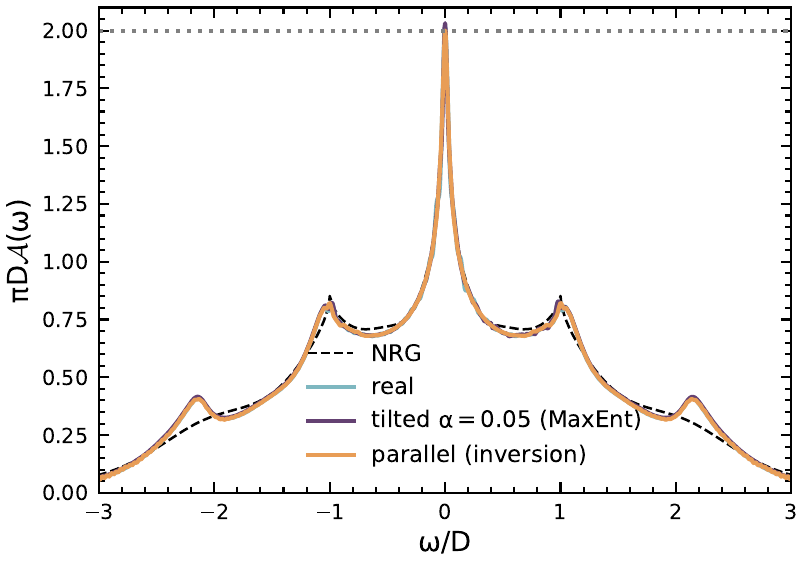}
  }%
  \\
  \subfloat[\label{fig:3band1}]{
    \hspace*{0.7em}
    \includegraphics[width=0.9\linewidth]{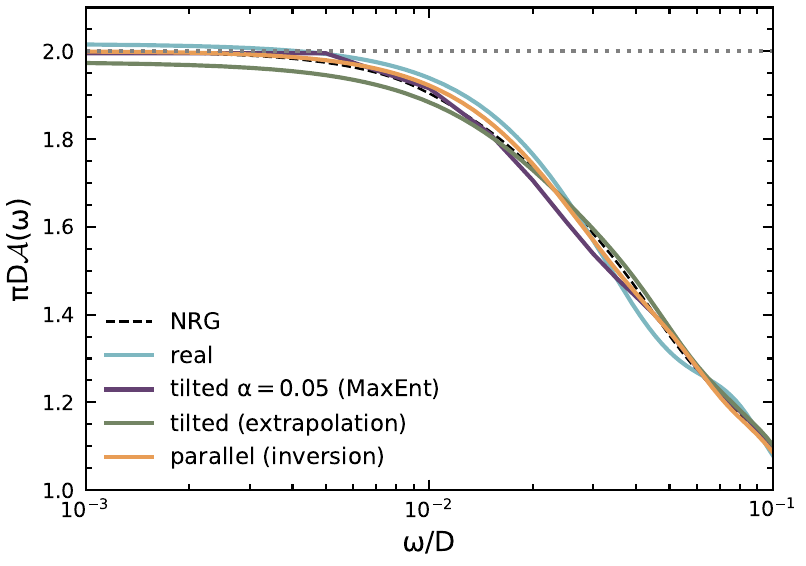}
  }
  \caption{
    \label{fig:HK_specs_combined}
     Spectral function $\mathcal{A}(\omega)$ for a three-orbital model with
     Hubbard--Kanamori interaction. All calculations are at truncated weight
     $w_t=10^{-10}$ and without broadening ($\eta=0$). Real-time results stem
     from a time evolution up to $t_{\mathrm{max}}=120D^{-1}$ with a maximum
     bond dimension $m=1024$. For the tilted (MaxEnt) method at angle $\alpha=0.05$ the corresponding values are
     $t_{\mathrm{max}}=180D^{-1}$ and $m=2048$, for the parallel (inversion) method at
     $\tau=1$, $t_{\mathrm{max}}=220D^{-1}$ and $m=1024$; in that case, an
     increased bath size $N_b=139$ was used. The post-processing for the parallel (inversion) method used $n=3$ contours at
     $\tau_k\in\{1, 1.15, 1.3\}$ with coefficients $a_k\in\{33.2, -57.8,
     25.6\}$ (rounded). The Friedel sum rule is obeyed to an accuracy of
     $0.2\%$ for the tilted (MaxEnt) data and $0.02\%$ for the parallel (inversion) data versus
     $0.8\%$ in the real-time calculation.
     ~\subfigref{fig:3band1} zoom into
     Fig.~\subfigref{fig:HK_global} with additional data obtained from the tilted (extrapolation) method using
     contours $\alpha \in [0.05,0.1,0.15,0.2,0.25,0.3]$,  averaging over all 4th order contributions. The
     oscillations in the real-time result are now clearly visible. The tilted (MaxEnt) result using $\alpha=0.05$ shows a similar
     unphysical slight dip between $\omega/D=10^{-2}$ and $10^{-1}$ as for the
     SIAM.
    }
\end{figure}
We again use NRG results as a benchmark, since it is currently the most accurate method available for low frequencies. 
The Hubbard--Kanamori Hamiltonian in its band-degenerate form has an SO(3) orbital symmetry which makes it accessible  
to standard multiorbital NRG \cite{Stadler2015,Horvat2016,Horvat2017,Kugler2019}
(without the need for interleaving the Wilson chain \cite{Mitchell2014,Stadler2016,Kugler2019,Kugler2020,Kugler2022prl}). 
Note that we do not exploit the SO(3) orbital symmetry in our T3NS calculations as it typically does not appear in multiorbital simulations of real materials.

\begin{figure}[h]
    \centering
    \includegraphics[width=0.9\linewidth]{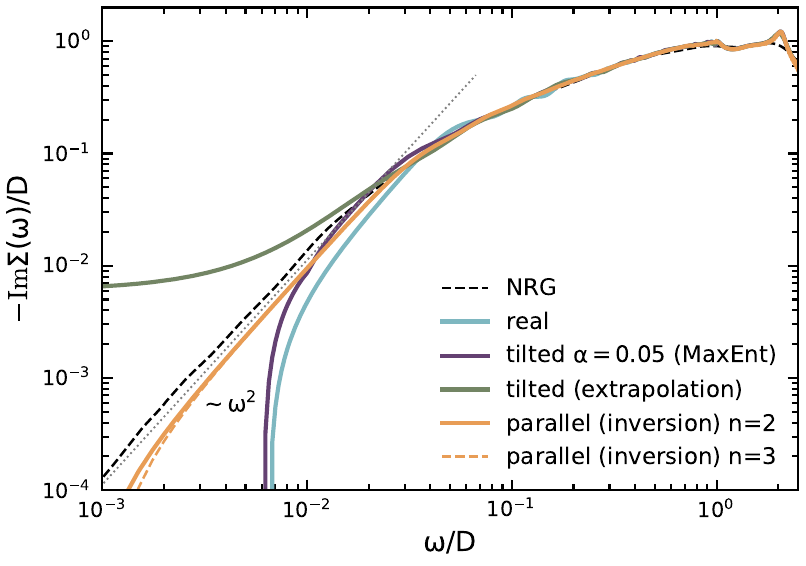}
    \caption{
       Imaginary part of the self-energy of the Hubbard--Kanamori
       Hamiltonian. The real-time result ("real") shows oscillations and
       fails at $\omega/D\approx 0.05$, missing the low-frequency physics.
       The dotted line indicates the $\omega^2$ Fermi-liquid behavior of the self energy.
       The tilted (MaxEnt) and tilted (extrapolation) methods do not reach this regime. 
       The parallel (inversion) method (with $n=2,3$ with $\tau_k\in\{1.15, 1.3\}$ and $a_k\in\{
       $8.7, -7.7$\}$) is in very good agreement with NRG data but
       for a prefactor (see text) down to $\omega/D\approx 0.002$. 
    }
    \label{fig:HK_SelfE_log}
\end{figure}

Despite the tailored representation on a tree-tensor network, pure real-time
calculations fail at high frequency resolution. The complex-contour calculations,
on the other hand, yield reliable results. 
In Figs.~\ref{fig:HK_global} and
\ref{fig:3band1} we show that the spectral function results from the tilted (MaxEnt) method and from the parallel
(inversion) method agree overall very well, whereas the real-time results show unphysical wiggles. 
There is however again a small discrepancy between the tilted (MaxEnt) and the parallel (inversion) methods at $\omega/D$ a bit above 0.01; in view of the results below we interpret this as an inaccuracy of the tilted (MaxEnt) result. NRG results agree excellently at low frequencies, but show the expected deviations for larger frequencies due to the logarithmic discretization of NRG.

We calculate the self-energy of the Hubbard--Kanamori Hamiltonian as in the
case of the SIAM, but add further methods to have some mutual benchmarking.
We obtain $G(\omega)$ in four different ways: (i) from a real-time
calculation; (ii) from a tilted (MaxEnt) calculation at $\alpha=0.05$; (iii) from a tilted (extrapolation) calculation;
and (iv) from a parallel (inversion) calculation at $\tau=1.3$. In cases (ii) and (iv), we obtain $\mathcal{A}(\omega)$ and then $\text{Im}\,
G(\omega)$
with the Kramers--Kronig transformation. 

In Fig.~\ref{fig:HK_SelfE_log}, we observe that the real-time result shows
weak, but unphysical oscillations at frequencies above $\omega/D\approx 0.05$
and fully misses the low-energy Fermi-liquid physics
$\mathrm{Im}\,\Sigma(\omega) \sim \omega^2$ at lower frequencies.
Both the
tilted (MaxEnt) and tilted (extrapolation) methods perform somewhat better, but also fail at low
frequencies. At higher frequencies, the tilted (extrapolation) result is in
perfect agreement with the parallel (inversion) result. 

\begin{figure}[h]
    \centering
    \includegraphics[width=0.9\linewidth]{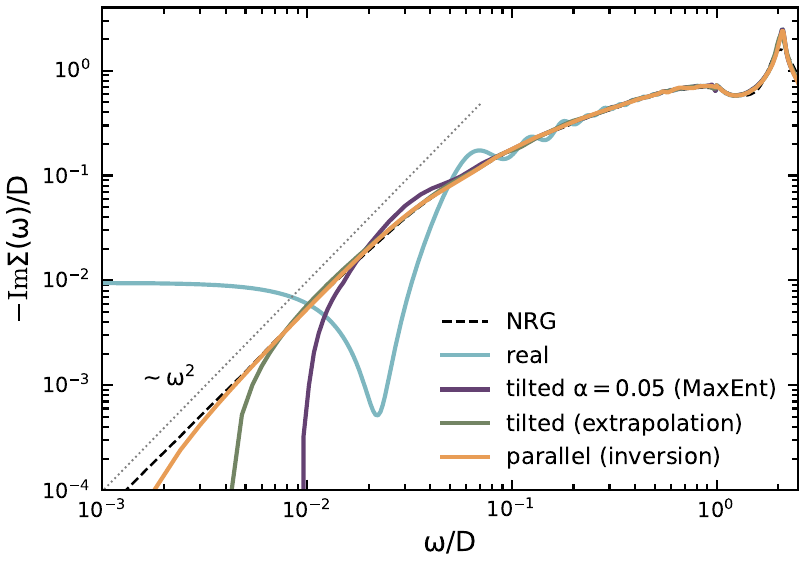}
    \caption{
       Imaginary part of the self-energy of the Dworin--Narath Hamiltonian. The
       parallel (inversion) method is in very good agreement with NRG data down to
       $\omega/D\approx 0.004$. All calculations are at truncated weight
       $w_t=10^{-10}$. Real-time evolution is up to
       $t_{\mathrm{max}}=120D^{-1}$ with a maximum bond dimension $m=512$. The
       tilted (MaxEnt) method uses angle $\alpha=0.05$ and goes to
       $t_{\mathrm{max}}=180D^{-1}$ and maximum bond dimension $m=1024$, the
       parallel (inversion) method uses $\tau=1$ and goes to $t_{\mathrm{max}}=140D^{-1}$ and
       maximum bond dimension $m=1500$, the bath size was increased to
       $N_b=139$. Results for the parallel (inversion) method are shown for $n=4$ with
       $\tau_k\in\{0.85, 1, 1.15, 1.3\}$; the coefficients $a_k$ (see main
       text) are then approximately $\{9.8, -14.2,  7.5,  -2.1\}$. 
 }
    \label{fig:DN_SelfE_log}
\end{figure}

The parallel (inversion) method easily
reaches the $\omega^2$-regime. As in the case of the SIAM, sum rule violations
in connection with the use of Dyson's equation lead to a breakdown at very low
frequencies. Interestingly, the breakdown does not occur at higher frequency
than in the SIAM case, indicating that this problem is not aggravated by the
higher complexity of the model studied.

In the case of the three-band Hubbard--Kanamori model, NRG provides the best results for comparison but is no longer an \textit{exact} benchmark: The large dimension of the local Hilbert space requires a rather large value of the discretization parameter, here $\Lambda=6$ \cite{Bulla2008}. Averaging over $n_z$ shifted discretization grids mitigates the effects of a coarse resolution of the hybridization function to some extent \cite{Zitko2009}. 
As discussed in Appendix~\ref{app:NRG_nz_extrap}, we found the best results by extrapolating to $n_z \to \infty.$
Generally, we observe that the NRG result for the $\omega^2$ coefficient of $-\Im\Sigma$ decreases with increasing resolution of the hybridization function (decreasing $\Lambda$, increasing $n_z$). It is therefore remarkable that the T3NS result is slightly
offset from the NRG result by a factor $< 1$ in the $\omega^2$ regime before the breakdown occurs at $\omega/D \approx 0.002$. 

We also considered the Dworin--Narath Hamiltonian, which can be represented similarly as in Eq.~\eqref{eq:Hubbard-Kanamori}, except that one necessarily has $U'=U-J$ and that the last term in Eq.~\eqref{eq:Hubbard-Kanamori} (known as pair hopping) is missing \cite{Georges2013}.
It obeys a larger orbital symmetry than the Hubbard--Kanamori Hamiltonian (SU(3) instead of SO(3)),
and thus allows for highly accurate NRG calculations at $\Lambda=4$ (see also Appendix~\ref{app:NRG_nz_extrap}). Figure~\ref{fig:DN_SelfE_log} shows the imaginary part of the self-energy for the Dworin--Narath Hamiltonian with the same parameters as before ($U/D=2$ and $J/D=0.3$). 
The performance of the time-evolution methods is similar to Fig.~\ref{fig:HK_SelfE_log}:
going toward low frequencies,
the real-time evolution soon yields unphysical results,
the tilted (MaxEnt) and tilted (extrapolation) schemes improve on this, 
while the best result by far is obtained via the parallel (inversion) method.
The corresponding curve follows the Fermi-liquid $\omega^2$ behavior down to $\omega/D \approx 0.004$.
Importantly, the agreement with NRG is excellent, as the coefficients of the
$\omega^2$ behavior match. This indicates that the small difference
in the prefactor of the self-energy in Fig.~\ref{fig:HK_SelfE_log} comes from
the fact that the Hubbard--Kanamori NRG result is not fully converged in all
numerical parameters. 

The parallel (inversion)
method emerges as the most reliable and performing among the methods tested here.
Moreover, it can be further improved systematically: it rests on an
analytically exact formula and the breakdown at very low frequencies occurs
because $G(t+i\tau)$ was not calculated for long enough times to yield highly
reliable $G_\tau(\omega)$ for very small $\omega$. As entanglement growth is
curtailed, the region of $\omega^2$ scaling can be extended to smaller
frequencies by longer time evolutions. Note that the complex-time evolutions
employed here easily generalize to less symmetric situations, as we did not use
any of the emerging larger symmetries, and also have sufficient numerical
efficiency to move to systems with more than three orbitals. Additionally, we
expect that the accuracy of the self-energies provided by complex time
evolution can be further improved by using improved estimators as in NRG
\cite{Kugler2022}.

\section{Conclusion}\label{conclusion}
Complex-time evolution is an addition to the toolbox of tensor network simulations that offers high resolution when computing Green's functions at low frequencies at a fraction of the computational cost. At the same time, it maintains the quality of high-frequency data previously available. The speedup is highest for the tilted contour, about two orders of magnitude. MaxEnt as a method of analytical continuation provides very good, but not excellent agreement with real-time data where the latter is essentially exact. 

The parallel contour calculations yielded the second highest speedup of (more than) an order of magnitude; post-processing by two different methods, inversion and extrapolation, provided excellent mutual agreement, allowing mutual control, as well as with benchmark data. Results were always somewhat more accurate than tilted  (MaxEnt) results. In the very low frequency regime for the self-energy of the three-band model, our most challenging calculation, the tilted (extrapolation) method ran into difficulties, but the parallel (inversion) method was stable and accurate: The calculation of the self-energy of the Hubbard--Kanamori and Dworin--Narath three-band models reaches successfully into the $\textrm{Im}\,\Sigma(\omega) \sim \omega^2$ Fermi liquid regime down to $\omega/D\approx 0.002$ in agreement with NRG benchmark results. 
The parallel (inversion) method therefore seems to provide the best compromise in speedup and accuracy. 
It can be systematically improved to reach even lower frequencies, albeit at mounting numerical cost; design decisions here will reflect compromises between required low-frequency resolution and available CPU time. 

The availability of multiple complex time-evolution schemes that can be
directly used or post-processed in different ways to extract real-time
information makes these methods very controlled. 
We expect these methods to be
particularly useful in the context of quantum-embedding methods using
frequency-based information such as DMFT and its derivatives, where efficiency
of impurity solvers is paramount. In fact, the very high efficiency of the
tilted contour calculations may make them the preferred approach for DMFT
despite the limited accuracy. It remains as a challenge for the future to apply
these methods in the context of global quenches or higher-dimensional systems
simulated directly on their real-space lattices. It is hoped that the
suppression of high-energy contributions by imaginary time evolution may give
access to long-time information inaccessible to real-time methods without
substantial loss of accuracy.  

\section{Acknowledgments}
We acknowledge useful discussions with Xiaodong Cao, Sebastian Paeckel and Miles Stoudenmire. M.G.\ and U.S.\ acknowledge support by the Deutsche Forschungsgemeinschaft (DFG, German Research Foundation) under Germany’s Excellence Strategy-426 EXC-2111-390814868. M.G.\ and U.S.\ thank the Flatiron Institute for its hospitality. The Flatiron Institute is a division of the Simons Foundation.

\appendix

\section{Proof of Eq.~\eqref{inversion_eq}}\label{app:AtauRelationProof}

Using the definition of the complex time evolution \eqref{def:ComplexTimeEvolution}
we obtain

\begin{align*}\label{defAtau}
   \expval{\{\hat{c}^{\nodagger}_{0}(z(t)),\hat{c}^{\dagger}_{0}\}}{\psi_0} =
\sum_A 
&|\langle \psi_0 | \hat{c}^{\nodagger}_{0} |A \rangle |^2 
 e^{-i E_A(t - i\tau)}
\\ +
& |\langle \psi_0 | \hat{c}^{\dagger}_{0} |A \rangle |^2 
 e^{+i E_A(t + i\tau)} 
\end{align*}
where $|A\rangle$ is a eigenstate basis of the many-body Hamiltonian $H - E_0$, and 
$E_A$ its eigenvalue.
From the definition of $\mathcal{A}_\tau$ \eqref{defAtau}, 
we have 
\begin{align*}
   \mathcal{A}_\tau(\omega)  =
\sum_A
&|\langle \psi_0 | \hat{c}^{\nodagger}_{0} |A \rangle |^2 
 \delta(\omega- E_A) e^{- E_A \tau}
+ \\
&|\langle \psi_0 | \hat{c}^{\dagger}_{0} |A \rangle |^2 
 \delta(\omega + E_A) e^{- E_A \tau}
\end{align*}
in the limit $\eta\rightarrow0$.
By definition of the ground state, $E_A>0$.
In the first term, $\omega = E_A = |\omega|$, while in the second term $E_A = - \omega = |\omega|$, 
so we get 
\begin{align*}
   \mathcal{A}_\tau(\omega)  =  e^{- |\omega| \tau}
\sum_A
 &|\langle \psi_0 | \hat{c}^{\nodagger}_{0} |A \rangle |^2 
 \delta(\omega- E_A) 
\ + \\ 
 &|\langle \psi_0 | \hat{c}^{\dagger}_{0} |A \rangle |^2 
 \delta(\omega + E_A)
\end{align*}
and therefore 
\begin{equation*}
 \mathcal{A}_\tau(\omega) =  \mathcal{A}_{\tau =0}(\omega)   e^{- |\omega| \tau}
\end{equation*}

\section{Extrapolation Method}\label{app:extrapolationmethod}
We assume that we know $n$ Green's functions $G(t+i\tau_k)$ for $n$ different $\tau_k$. As seen previously, they can be generated at low numerical cost from a single contour. We approximate the behavior by a power series incorporating terms up to order $n-1$,
\begin{equation}
    G(t+i\tau) = G(t) + \sum_{m=1}^{n-1} \tau^m h_m(t).
\end{equation}
For $n$ values $\tau_k$, we have $n$ equations with $n$ unknown variables, the Green's function $G(t)$ we are interested in and $n-1$ coefficients $h_m(t)$ which we will not require explicitly. This linear equation system can be written as ${\bf g} = {\bf M} \cdot {\bf h}$, where the vector components are $g_k = G(t+i\tau_k)$, $h_0=G(t)$ and $h_k = h_k(t)$ for $k>0$, and 
\begin{equation}
M_{kl} = \tau_k^{l-1} .
\end{equation}
An inversion of ${\bf M}$ yields
\begin{equation}
    G(t) = \sum_{k=1}^{n} (M_{1\, k})^{-1} \, G(t+i\tau_k).
\end{equation}
For increasing $n$ the inversion of ${\bf M}$ becomes more difficult. To stabilize it, it is useful to re-scale both $h_k(t)$ and ${\bf M}$ such that  $G(t)$ remains unaffected. With $\tau_{\textrm{max}}$ the largest $\tau_k$, we re-scale as
\begin{equation}
    M_{kl} = \Big{(}\frac{\tau_k}{\tau_{\textrm{max}} }\Big{)}^{l-1}.
\end{equation}
If we want to solve at order $n$, but know $G(t+i\tau_k)$ for more than $n$ different $\tau_k$, we found a substantial increase of accuracy by averaging over all different $n$th-order extrapolations with different choices of $n$ values of $\tau_k$. In practice, we used $n=6$ for 13 different $G(t+i\tau_k)$. In the calculations shown here (Figs.~\ref{fig_inversion} and \ref{parallel:zoom}), they were equidistantly ($\delta \tau=0.075$) centered on $\tau=1$. (Note that the extrapolation method can also be applied to the linear contours of Sec.~\ref{sec:inclined}.)
\\
\\
\section{Analytic Continuation}\label{ac_sec}
For the analytical continuation by MaxEnt, we follow the procedure outlined in \cite{Guther2018}.
There is a connection between $G(z)$ and the spectral function $\mathcal{A}$ in the case of complex\hyp time contours, given by \begin{equation}
    G\big{(} z \big{)} = -i \Theta(t) \int_{-\infty}^{\infty}d\omega \mathcal{A}(\omega) K\big{(}z, \omega\big{)},
\end{equation}
where the integration kernel $K(z,\omega)$ is defined as \begin{equation*}
    K(z,\omega) = 
    \begin{cases}
        \exp\big{(}-i\overline{z}\,\omega\big{)}, \quad \text{if} \;\; \omega \geq 0 \\
        \exp\big{(}-i z\,\omega\big{)}, \quad \text{if} \;\; \omega < 0 
    \end{cases}
\end{equation*}
As the kernel may be complex valued, the equations have to be slightly adapted \cite{Guther2018}. Besides this, the well established methods of MaxEnt were used \cite{jarrell1996bayesian, skilling1989classic}.
We found analytic continuation to be rather stable with this kernel. However, care must be taken at low frequencies as the kink of the kernel at zero frequency leads to results deviating from spectral functions obtained from purely real\hyp time evolution. We propose two different measures as possible resolutions. First, we use a broadening around the discontinuity, by replacing the problematic $\text{sgn}(\omega)$ with $\tanh(\nicefrac{\omega}{\sigma})$, with the free factor $\sigma$. The kernel then reads
\begin{equation}
    K(z,\omega) = \exp\big{(}-(it + \tanh(\nicefrac{\omega}{\sigma})\tau)\omega\big{)}
\end{equation}
In the limit $\sigma \to 0$ this kernel is exact. Hence, $\sigma$ has to be chosen sufficiently small to leave the overall structure of the inversion problem unchanged. We found sensible values to be smaller than the width of the main peak, at $\sigma = 0.11$. Although this change helps a bit, we found it to be not very reliable, leading to strong oscillations around $\omega=0$. We therefore implemented a second method, which introduces a correction term to $Q(\mathcal{A}) = -\chi^2+S(\mathcal{A})$, maximized in normal MaxEnt. We rather take
\begin{equation}
     Q(\mathcal{A}) = -\chi^2 + \alpha S(\mathcal{A}) - \beta \int_{-\infty}^{\infty} d\omega \Big{(}\frac{d\mathcal{A}}{d\omega}\Big{)}^2 f(\omega).
\end{equation}
The additional last term favors smooth spectral functions, since, when maximizing $Q$, it tries to minimize the quadratic slope of $\mathcal{A}$. The observed quick oscillations around zero are thus smoothed out, since they increase the quadratic slope. $f(\omega)$ serves as a weighting factor focusing on small frequencies. We used a Gaussian $f(\omega) = \exp(-(\nicefrac{\omega}{\sigma})^2)$. The new parameter $\beta$ controls this correction term; $\beta = \alpha$ turned out to be a viable choice. For all MaxEnt calculations involving the three-band model, we chose $\sigma=1$. The MaxEnt with correction term maintains the correct peak heights. We leave an even better stabilization of the Kernel at low frequencies to further research. 

\section{Real Frequency Results Without Analytic Continuation: Kink contour}\label{real_w_ac}
\begin{figure}
  \centering
  \subfloat[\label{fig:contour_kink}]{
    \includegraphics[width=0.65\linewidth]{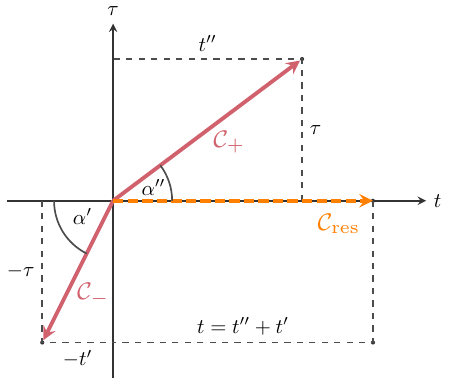}
  }%
  \\
  \subfloat[\label{fig:contour_kink_stable}]{
    \includegraphics[width=0.65\linewidth]{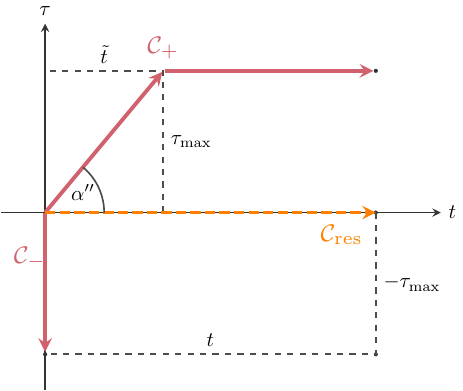}
  }
  \caption{
    \label{fig:kink_contours}
    \subfigref{fig:contour_kink} Contour for complex time evolution with no postprocessing.
    \subfigref{fig:contour_kink_stable} Stable version with a horizontal contour, Cf Text. 
    }
\end{figure}

For an additional approach, we reformulate the idea of time splitting in the complex plane in a different way by splitting {\em real} time into two {\em complex} times via $t=z'+z''$ with $z'=t'-i\tau$, $z''=t''+i\tau$, $t, t', t'',\tau >0$ where $t=t'+t''$. Then 
\begin{equation}
\bra{\psi_0} c^\dagger (0) c(t) \ket{\psi_0} =  [c(-\overline{z}') \ket{\psi_0}]^\dagger c(z'') \ket{\psi_0}
\end{equation}
Note that in this case no analytical continuation or any other post-processing is required. If we continue to use linear contours in the complex plane, we get two contours $\mathcal{C_+}$ and $\mathcal{C_-}$ in the upper right and lower left quadrants of the complex plane at angles $\alpha'$ and $\alpha''$ which need not be identical as no symmetric splitting of $t$ is required (see  \cref{fig:contour_kink}). The two angles $\alpha', \alpha''$ can be chosen freely, but the imaginary time steps in $\mathcal{C_+}$ suppress the entanglement growth of the real-time evolution, whereas the same steps increase that entanglement growth on contour $\mathcal{C_-}$. This suggests a splitting where $\alpha'>\alpha''$ and hence $t' < t''$. In our calculations, we chose $\alpha''=0.1$ and $\alpha'=\pi/2$, i.e. a purely imaginary time evolution on a vertical contour  $\mathcal{C_-}$; $t=t''$ and $t'=0$. 
Numerical instabilities in the time evolution on $\mathcal{C_-}$ limit our maximum evolution time on the corresponding contour $\mathcal{C_+}$ to about $|z''|=20D^{-1}$ or $\tau''=1.996 D^{-1}$ for $\alpha''=0.1$ and $\alpha'=\pi/2$. 
\begin{figure}[h]
  \centering
  \subfloat[\label{comp_Gret_imag}]{
    \includegraphics[width=0.85\linewidth]{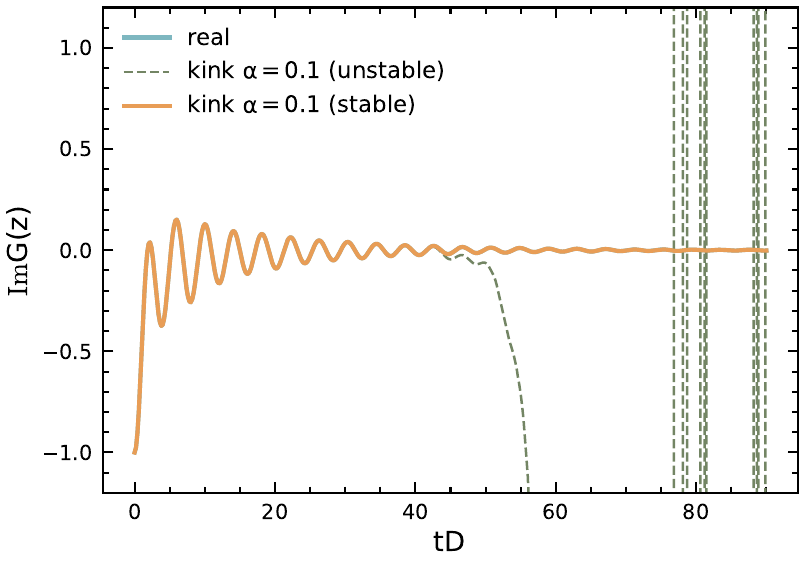}
  }%
  \\
  \subfloat[\label{fig:kinkmethod}]{
      \hspace*{-0.6em}
    \includegraphics[width=0.86\linewidth]{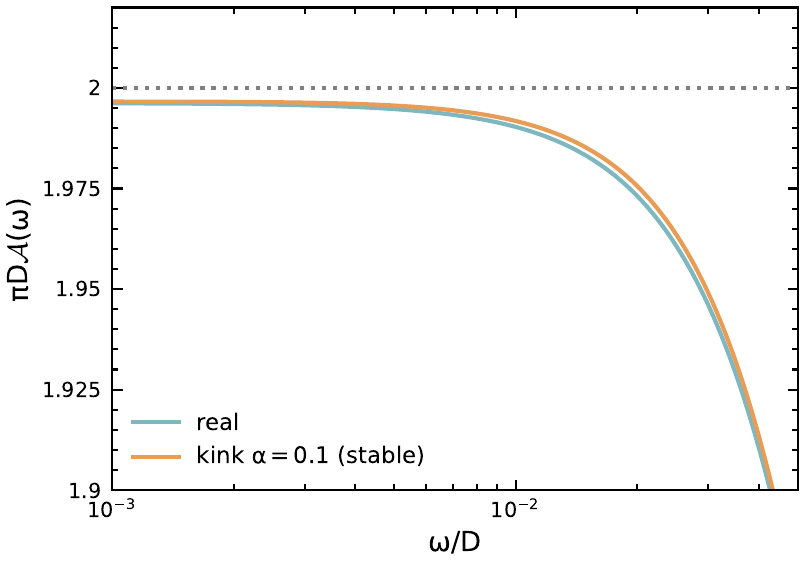}
  }
  \caption{
    \label{fig:kink_imag_full}
Imaginary part of the retarded Green's function vs $t$ and spectral function vs $\omega$ for the SIAM at $U/D=2$ and
 other parameters as before, obtained using the kink contour method with
$\alpha''=0.1$ and $\alpha'=\frac{\pi}{2}$. 
The "stable" label refers to the second contour in Fig.~\ref{fig:contour_kink}
with $t_{\mathrm{max}}=20D^{-1}$ or $\tau_{\mathrm{max}} =0.998=1.996D^{-1}$.
The "stable" contour result is in excellent agreement with the real time reference data.
}
\end{figure}
The numerical instabilities in the time evolution on $\mathcal{C_-}$ can be seen clearly in \cref{comp_Gret_imag}. It might be suspected that the exponential growth of the weight of higher-energy states on contour $\mathcal{C_-}$ suppresses the contribution from low-energy states which are enhanced on contour $\mathcal{C_+}$ and numerical cancellations fail, leading to the instabilities. 
At the onset of the instabilities, the norm difference is only of the order $10^4$, which should only lose insignificant digits. So even though exponential growth will eventually make this method unstable, it does not seem to be the origin of the currently observed instabilities. We suspect that they are related to the relatively subtle interplay of errors in the TDVP method which suggests that improvements are possible at the level of the time-evolution method. 

The instability is mended by continuing with a contour parallel to the real axis for larger times, before instabilities occur, see Figure~\ref{fig:contour_kink_stable}. Figure~\ref{fig:kinkmethod} shows the excellent quality of the result for the spectral function of the SIAM.

\begin{figure}[h]
\vspace{0.5cm}
    \centering
    \includegraphics[width=0.85\linewidth]{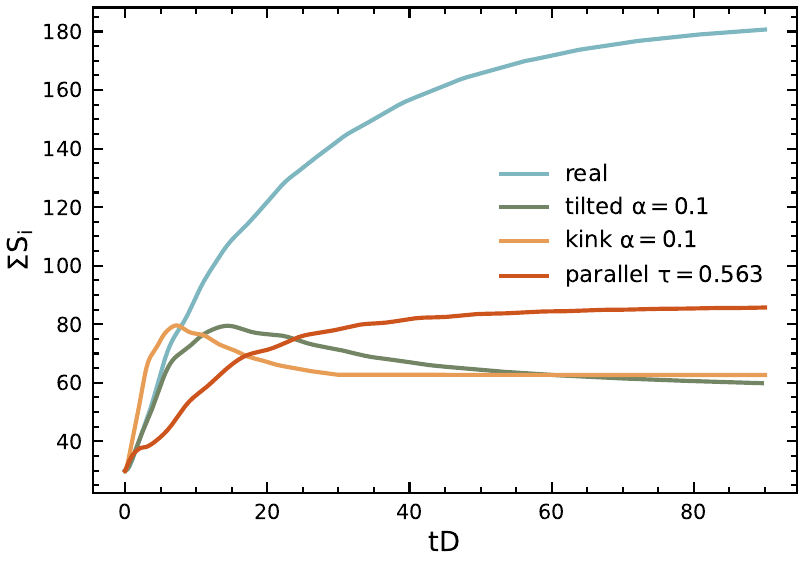}
    \caption{Comparison of the entanglement entropy on different complex\hyp time contours. As the system is heterogeneous, the sum of the entanglement entropies for all system cuts is shown. Kink $\alpha=0.1$ is the post-processing free contour with vertical contour along the negative imaginary axis and a tilted contour in positive imaginary direction until $t=30D^{-1}$) and $\tau=0.563$ a parallel contour with constant imaginary part $\tau=0.563$. Times of contours exploiting real\hyp time splitting were multiplied by a factor 2 to represent the same final time.}
    \label{comp_contours}
\end{figure}

This method is at the moment the least performant in speed, roughly a speed\hyp up of a factor three, but attractive due to the absence of post-processing. Methodological progress in the time-evolution methods may make it fully competitive. 

\section{Comparison of Contours}\label{app:comp_contours}

The ideal choice of the complex contour is ultimately reliant on the model and
its specific high energy quasi\hyp particle peak structure, because the loss of
high-energy information is greater if the contour is further away from the
real\hyp time axis during early times. A second crucial aspect is the growth of
entanglement entropies on the complex contours, as it
determines the computational cost. This is illustrated on \cref{comp_contours}.
Not surprisingly, the real-time contour has
the strongest overall growth of entanglement, which does not saturate,
explaining the vastly larger numerical resources required. The tilted contour
(here $\alpha=0.1$) starts with a similar growth of entanglement at early times
when it is close to the real axis, but actually reveals even a slight decrease
before saturation at some relatively small value. The kink contour without
post-processing (kink $\alpha=0.1$) has a very strong initial growth of
entanglement, even stronger than for real-time evolution, because it does not
exploit time\hyp splitting in the real\hyp time direction. Ultimately, though,
entanglement strongly resembles that of the tilted contour, reminiscent of the
observed saturation of entanglement growth for all complex\hyp time contours.
Finally, the parallel contour at constant $\tau$ has the slowest early growth
of entanglement as it aggressively suppresses high-energy states, but then
settles on a somewhat larger saturation value as the contour stays closer to
the real axis. (Note that the $\alpha=0.1$ and $\tau=0.563$ data do not allow
for immediate comparison; they are rather indicative of typical behavior for
the respective contours.)  The saturation of entanglement observed for all
complex contours explains the large accessible times.

\section{NRG extrapolation in $n_z$}\label{app:NRG_nz_extrap}

For the cases studied in this paper, the self-energy obeys the zero-temperature Fermi-liquid behavior $-\Im\Sigma(\omega) = c \, \omega^2$ at low frequencies $\omega$.
The NRG result for $c$ is sensitive to the discretization of the hybridization function, i.e., it varies notably with the discretization parameter $\Lambda$ \cite{Bulla2008} and the number of $z$-shifts $n_z$ \cite{Zitko2009}.
The hypothetical continuum limit $\Lambda \to 1$ is numerically not feasible. For one-orbital calculations, $\Lambda$ can be chosen as small as, say, $1.7$; three-orbital calculations require $\Lambda \gtrsim 4$.
The limit of $n_z \to \infty$ similarly is not feasible. For our one-orbital calculations, we consider $n_z$ values up to $12$; for our three-orbital calculations, $n_z \leq 6$.

For the particle-hole symmetric Anderson model with a flat hybridization, 
$-\Im\Delta(\omega)=\Delta \Theta(D-|\omega|)$,
in the wide-band limit ($D \to \infty$) and at weak interaction
$u \equiv U/(\pi\Delta) \ll 1$,
the coefficient $c$ is known from second-order perturbation theory \cite{Zlatic1983}.
This applies to the one-orbital case \cite{Yamada1975} and can be readily generalized to multiple orbitals \cite{Nishikawa2010} (setting $J=0$ for simplicity).
The result, with $M$ the number of orbitals, is
\begin{equation}
c_{\mathrm{exact}} = \tfrac{1}{2} u^2 M .
\end{equation}

\begin{figure}[t]
\vspace{0.5cm}
    \centering
    \includegraphics[width=0.77\linewidth]{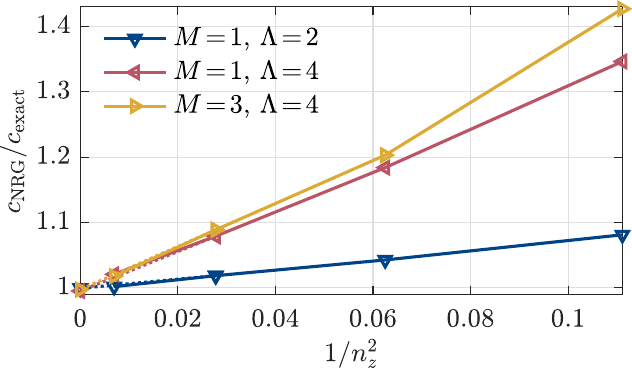}
    \caption{Coefficient $c$ of $-\Im\Sigma(\omega)=c \, \omega^2$ obtained in NRG calculations at weak interaction for various $M$, $\Lambda$, and $n_z \in \{3,4,6,12\}$. For each $M$ and $\Lambda$, one observes a roughly linear dependence of $c_{\mathrm{NRG}}$ in $1/n_z^2$, permitting an extrapolation toward the ideal $n_z=\infty$ limit. The extrapolated value is in perfect agreement with the analytically known result $c_{\mathrm{exact}}$.}
    \label{NRG_nz_extrap}
\end{figure}

We computed NRG results using $D=100$, $\Delta=1$, $U=0.2$, $T=0$.
The coefficient $c$ is deduced from a fit in the in the frequency range $5 \cdot 10^{-4} < \omega < 5 \cdot 10^{-2}$,
after adaptive broadening \cite{Lee2016} with $\alpha_z = \alpha / n_z$ and $\alpha = 1$.
Figure~\ref{NRG_nz_extrap} shows $c_{\mathrm{NRG}}$, for one- and three-orbital calculations and selected values of $\Lambda$, for various values of $n_z$.
One observes a roughly linear behavior in $1/n_z^2$, nicely extrapolating to the analytic value $c_{\mathrm{exact}}$ at $n_z = \infty$.
Not surprisingly, the dependence of $c_{\mathrm{NRG}}$ on $n_z$ is stronger for larger $\Lambda$.
We determined the extrapolated values from the two data points at $n_z=4$ and $n_z=6$.
The same choice is used for the frequency-dependent self-energies extrapolated in $n_z$ shown in the main text.

\begin{table}[t]
\centering
\begin{tabular}{c|c|c|c|c}
\#orbitals $M$ & $H_{\mathrm{int}}$ & Symmetry & $N_{\mathrm{kp}}$ & $N_{\mathrm{kp}}^*$
\\ \hline
1 & -
& 
$\textrm{SU}(2)_{\textrm{ch}} \otimes \textrm{SU}(2)_{\textrm{sp}} $
& 
$32\textsf{k}$ & $300\textsf{k}$
\\ \hline
3 & DN
& 
$\textrm{U}(1)_{\textrm{ch}} \!\otimes\! \textrm{SU}(2)_{\textrm{sp}} \!\otimes\! \textrm{SU}(3)_{\textrm{orb}}$
& 
$20\textsf{k}$ & $800\textsf{k}$
\\ \hline
3 & HK
& 
$\textrm{U}(1)_{\textrm{ch}} \!\otimes\! \textrm{SU}(2)_{\textrm{sp}} \!\otimes\! \textrm{SO}(3)_{\textrm{orb}}$
& 
$15\textsf{k}$ & $200\textsf{k}$
\end{tabular}
\caption{Symmetries used and number of multiples $N_{\mathrm{kp}}$ (or approximate number of states $N_{\mathrm{kp}}^*$) kept during the NRG iterative diagonalization ($1\textsf{k}$ denotes $10^3$).}
\label{tab:NRG_setting}
\end{table}

Table~\ref{tab:NRG_setting} summarizes the symmetries used in NRG (\textbf{ch}arge, \textbf{sp}in, and \textbf{orb}ital symmetries) and the number of states kept during the iterative diagonalization. In the three-orbital case, one can distinguish the Hubbard--Kanamori (HK) and the Dworin--Narath (DN) Hamiltonians \cite{Georges2013}, with SO(3) and SU(3) orbital symmetry, respectively.
The large SU(3) orbital symmetry of the DN Hamiltonian (also used here at $J=0$) allows us to effectively keep a much larger number of states $N_{\mathrm{kp}}^*$ than in the calculations of the HK Hamiltonian, so that the former results can be considered significantly more accurate than the latter.

\bibliography{literature}
\end{document}